\documentclass[times,review,preprint]{elsarticle}

\usepackage{jcomp}
\usepackage{lineno}
\usepackage{framed,multirow}

\usepackage[fleqn]{amsmath}
\usepackage{amssymb,amsbsy}
\usepackage{latexsym}

\newcommand{\ddx}[2]{\frac{\partial #1}{\partial #2}}

\usepackage{bm}
\newcommand{\myvec}[1]{\bm{\mathrm{#1}}}
\newcommand{\bmu}{\myvec{u}}
\newcommand{\bmx}{\myvec{x}}
\newcommand{\Sext}{S^{\mathrm{ext}}}
\newcommand{\CFL}{\mathit{CFL}}
\newcommand{\mybar}{\rule[-0.4ex]{0.2ex}{1.0em}}

\usepackage{pifont}
\usepackage[dvipsnames]{xcolor}
\definecolor{newcolor}{rgb}{.8,.349,.1}
\usepackage[pdfstartview=FitH,bookmarks=false,bookmarksopen=false]{hyperref}

\journal{}%Journal of Computational Physics}

\begin{document}

\verso{M. A. Hossain, S. Pimentel and J. M. Stockie}

\begin{frontmatter}

  \title{Simulating surface height and terminus position
    for marine outlet glaciers using a level set method with data
    assimilation}%\tnoteref{tnote1}}%
%%
%\title{Using data assimilation and the level set method to model surface and terminus change in marine outlet glaciers \tnoteref{tnote1}}%
%
%\title{Data assimilation using the level set method to model surface and terminus change in marine outlet glaciers \tnoteref{tnote1}}%
%\title{Using data assimilation and the level set method to model marine outlet glaciers \tnoteref{tnote1}}%
%\title{Using data assimilation and the level set method to model marine outlet glacier dynamics \tnoteref{tnote1}}%
%\title{Modelling marine outlet glacier dynamics using data assimilation and the level set method \tnoteref{tnote1}}%
%Assimilation of terminus change and surface observations in marine outlet glaciers using level set method \tnoteref{tnote1}}%

%\tnotetext[tnote1]{Preprint submitted to Elsevier.}

%\author[1]{M. Alamgir \snm{Hossain}\corref{cor1}}
\author[1]{M. Alamgir Hossain\corref{cor1}}
\cortext[cor1]{Corresponding author: 
  Tel.: +1-604-779-7064;  
  fax: +1-778-782-4947;}
\ead{mahossai@sfu.ca}

%\author[2]{Sam \snm{Pimentel}}
\author[2]{Sam Pimentel}
\ead{sam.pimentel@twu.ca}
%\author[2]{Sam \snm{Pimentel}\fnref{fn1}}
%\fntext[fn1]{This is author footnote for second author.}  

%\author[1]{John M. \snm{Stockie}}
\author[1]{John M. Stockie}
\ead{jstockie@sfu.ca}
%% Third author's email
%\author[1]{M. Alamgir \snm{Hossain}}

\address[1]{Department of Mathematics, Simon Fraser University, Burnaby,
  BC, Canada} 
\address[2]{Department of Mathematical Sciences, Trinity Western
  University, Langley, BC, Canada} 

\received{1 Jan 2022}
\finalform{1 Jan 2022}
\accepted{1 Jan 2022}
\availableonline{1 Jan 2022}
\communicated{N. NNNN}

\begin{abstract}
%%%%
  We implement a data assimilation framework for integrating ice surface
  and terminus position observations into a numerical ice-flow model.
  The model uses the well-known shallow shelf approximation (SSA)
  coupled to a level set method to capture ice motion and changes in the
  glacier geometry.  The level set method explicitly tracks the evolving
  ice--atmosphere and ice--ocean boundaries for a marine outlet glacier.
  We use an Ensemble Transform Kalman Filter to assimilate observations
  of ice surface elevation and lateral ice extent by updating the level
  set function that describes the ice interface. Numerical experiments
  on an idealized marine-terminating glacier demonstrate the
  effectiveness of our data assimilation approach for tracking seasonal
  and multi-year glacier advance and retreat cycles. The model is also
  applied to simulate Helheim Glacier, a major tidewater-terminating
  glacier of the Greenland Ice Sheet that has experienced a recent
  history of rapid retreat. By assimilating observations from
  remotely-sensed surface elevation profiles we are able to more
  accurately track the migrating glacier terminus and glacier surface
  changes. These results support the use of data assimilation
  methodologies for obtaining more accurate predictions of short-term
  ice sheet dynamics.
%%%%
\end{abstract}


\end{frontmatter}

%\linenumbers


%%%%%%%%%%%%%%%%%%%%%%%%%%%%%%%%%%%%%%%%%%%%%%%%%%%%%%%%%%%%%%%%%%
%%%%%%%%%%%%%%%%%%%%%%%%%%%%%%%%%%%%%%%%%%%%%%%%%%%%%%%%%%%%%%%%%%
\section{Introduction}

The rapid increases in ice loss from the Earth's cryosphere over the
past few decades~\cite{Slater_etal_2021} and ongoing uncertainty
regarding the future of the ice sheets \cite{catania2020future,
  pattyn2020uncertain} highlight the urgent need to integrate any
available in-situ and remotely-sensed observations along with
state-of-the-art numerical models \cite{kirchner-etal-2011} in order to better understand current
and future changes.  For instance, ice dynamical imbalances have been a
significant source of ice loss with increased discharge from outlet
glaciers in both Antarctica \cite{Mouginot_etal_2014} and the Greenland
Ice Sheet \cite{howat_etal_2008}.  Tidewater outlet glaciers are highly
sensitive to changes occurring at their terminus
\cite{catania2020future, king2020dynamic}, with potentially dramatic
consequences \cite{Edwards_etal_2019, Robel_etal_2019,
  Williams_etal_2021} that make it challenging to confidently predict
the rate of future mass loss.  Because of these dynamic instabilities,
considerable uncertainty remains when projecting dynamic ice behaviour
into the future.  Predicting short-term (inter- and intra-annual,
multi-year) ice dynamics could be improved with the seamless integration
of time-ordered observations into dynamical ice sheet models.  In this
paper, we develop a data assimilation approach combined with a level set
method to estimate the ice surface and terminus position over seasonal
and multi-year time scales.

Data assimilation combines observational data with a dynamical system
model to compute accurate estimates of the current and future states of
the system together with a measure of the uncertainty in those
states~\cite{law2015data}. 
%One of the main limitations in forecasting future sea level is uncertainties in the predictions of mass loss from the Greenland and Antarctic ice sheets~\cite{king2020dynamic,  solomon2007climate}. 
Data assimilation can be helpful in reducing
uncertainties in model state and
parameters~\cite{parrish2012toward}. Historically, data assimilation has
been used in operational meteorology and
oceanography~\cite{kalnay2003atmospheric}, while more recently it
has been applied to subjects as diverse as seismology~\cite{hoshiba2015numerical}, nuclear
fusion~\cite{morishita2020data}, medicine~\cite{cogan2021data} and
agronomy~\cite{jin2018review}.
% Direct measurements and satellite remote sensing of ice sheets and
% glaciers are increasing.  
Glaciologists have also begun exploiting data assimilation techniques to
incorporate time-ordered observations into their modelling studies
\cite{bonan2017data, bonan2014etkf, gillet2020assimilation}, and with
the recent growth in availability of observations
\cite{DCheng_etal_2021, Friedl_etal_2020, lenaerts2019observing,
  Tedesco_2014} it is only natural that data assimilation will increase
in importance in the glaciology community.

Due to the paucity of observations measuring glacier bed properties ice
sheet modellers have long employed static inverse techniques to determine
basal conditions from surface velocity observations
\cite{goldberg2011data, MacAyeal_1992, Petra_etal_2012}.  These methods
have also been used for estimating other unknowns, such as optimal
initial conditions from available observations~\cite{perego2014optimal}.
More recently, time-dependent adjoints have been considered for state
and parameter estimation as well as sensitivity analysis
\cite{Cheng_etal_2021, Goldberg_Heimbach_2013}.  In this paper, we
utilize a sequential data assimilation method based on the ensemble
Kalman filter (EnKF), which is a Monte Carlo approach to the Bayesian
update problem.  Such methods have also been applied in glaciology.  For
example, Bonan et al.~\cite{bonan2014etkf} developed an Ensemble
Transform Kalman Filter (ETKF) \cite{bishop2001adaptive,
  hunt2007efficient} for a Shallow Ice Approximation (SIA) model of ice
dynamics in order to estimate jointly the bedrock topography, ice
thickness and basal sliding parameter for an ice sheet. Bonan et
al.~\cite{bonan2017data} further investigated such data assimilation
approaches to estimate the state of a grounded ice sheet simulated using
a moving mesh method. Gillet-Chaulet~\cite{gillet2020assimilation} has
recently published a study of the EnKF method to initialize a marine ice
sheet model including grounding line migration.  There is immense
potential to further exploit these sequential data assimilation
techniques in forecast models of marine ice sheets, tidewater glaciers
and rapidly retreating outlet glaciers.

The level set method (LSM) is a versatile numerical technique for
simulating the motion of dynamically evolving surfaces and has found
many applications in image processing~\cite{osher2003geometric}, fire
front propagation~\cite{mallet2009modeling, sethian1999level}, tumor
growth~\cite{macklin2006improved, martins2007multiscale}, materials
science~\cite{nanthakumar2016detection}, computational fluid
dynamics~\cite{sethian1999level}, among others. The LSM was first
employed to track ice surfaces by Pralong and
Funk~\cite{pralong2004level}, who simulated the steady-state
configuration of a glacier by computing the stationary form of the level
set equation.
% coupled with the ice-air flow problem. This study did not
% consider the ice-water interface. 
Pralong and Funk~\cite{pralong2004level} compared different numerical
solutions of the model with other published analytical and numerical
solutions. Bondzio et~al. \cite{bondzio2016modelling} have also
developed a LSM-based 2D plan-view model to study the dynamic evolution
of a glacier calving front, but they did not consider the elevation of
the ice sheet. Recently, we have developed a level set algorithm for
explicitly tracking ice interfaces, moving ice margin, and grounding
line position in the context of a fixed grid finite-difference
scheme~\cite{hossain2020modelling}, which challenges the assertion
in~\cite{vieli-payne-2005} that fixed grid methods are inadequate for
capturing grounding line or terminus position.

There have been some recent attempts at implementing data assimilation
for applications using the level set method. One example is front
tracking for oil spills, for which Li et al.~\cite{li2017level} proposed
a level set based image assimilation method that captures the complex
topological structure of oil slicks and polluted regions to minimize any
delays in cleanup efforts because of inaccurate predictions from
observations.  A second example relates to wildfire spread, where Mandel
et al.~\cite{mandel2009data} used an EnKF method to estimate the state
of a wildland fire using available data and a model which accounts for
interactions between the fire and the atmosphere. These authors
implemented a level set method coupled with an atmospheric model to
deliver short-term predictions of fire propagation in a timely and
expedient manner.

We develop a glacier model that estimates ice surface elevation and terminus
position by combining the level set approach developed by Hossain  et
al.~\cite{hossain2020modelling} with the advantages of data
assimilation when observations are available to reinitialize the model.   
%We develop an ensemble transform Kalman filter (ETKF) to estimate the
%ice interface and grounding line position within annual scale magnitude
%by our level set method (Hossain  et al.~\cite{hossain2020modelling})
%for the initialisation of a marine ice sheet model.   
We begin in Section~\ref{sec:icefloweq} by introducing our ice sheet
model, which is a shallow shelf approximation (SSA)
\cite{bueler2009shallow} that is closely related to the SIA model.
Section~\ref{sec:lsm} describes the details of the level set method used
to capture the evolving ice sheet boundary. In Section~\ref{sec:da} we
apply the Ensemble Transform Kalman Filter (ETKF) to our state
estimation problem, and in Section~\ref{sec:da_algorithm} we outline our
hybrid SSA--LSM--ETKF algorithm that combines the various algorithm
components. In Section~\ref{sec:dataassimilation_results} we demonstrate
the effectiveness of the SSA--LSM--ETKF method for simulating an
idealized ice sheet test problem, as well as a more realistic problem
using actual data taken from Helheim Glacier, Greenland.

%%%%%%%%%%%%%%%%%%%%%%%%%%%%%%%%%%%%%%%%%%%%%%%%%%%%%%%%%%%%%%%%%%
%%%%%%%%%%%%%%%%%%%%%%%%%%%%%%%%%%%%%%%%%%%%%%%%%%%%%%%%%%%%%%%%%%
\section{Ice flow equations}
\label{sec:icefloweq}

In this section, we introduce the ice sheet model corresponding to the
shallow shelf approximation (SSA) for ice streams. We also describe our
numerical approach for computing the SSA velocities, which are used as
input to the LSM algorithm for evolving the ice sheet and tracking the
ice--air and ice--water interfaces (described later in
Section~\ref{sec:lsm}). The ETKF algorithm is then used to update the
ice surface and terminus positions using observations (see
Section~\ref{sec:da}).

The flow of an ice sheet is a large-scale fluid dynamics problem that
has many features in common with other geophysical systems such as
atmospheric or ocean circulation.  As such we describe the ice motion
using a velocity field $\bmu(x,y,z,t)$ that is governed by the
Navier-Stokes equations:
\begin{linenomath}
\begin{align}
  \frac{\partial \rho}{\partial t} + \nabla \cdot (\rho \bmu) &=
  0, \label{eq:compressible}\\ 
  \rho(\bmu_t + \bmu\cdot\nabla \bmu) &= -\nabla p + \nabla\cdot
  \tau_{i,j} + \rho \myvec{g}, \label{stokes1} 
\end{align}
\end{linenomath}
where $\rho$ is the ice density, $p$ is pressure, $\tau_{i,j}$ is the
deviatoric stress tensor, and $\myvec{g} = (0, 0, -g)$ is the
acceleration due to gravity. The uppermost compressible layer of an ice
sheet is shallow compared to the entire body of
ice~\cite{cuffey2010physics}, and consists of snow and firn (compacted
snow that is transforming into glacier ice). Moreover, there is little
variation in ice density due to changes in depth or temperature, so that
the ice can be considered an incompressible fluid for which the mass
conservation equation~\eqref{eq:compressible} simplifies to
$\nabla \cdot \bmu = 0$.

%%%%%%%%%%%%%%%%%%%%%%%%%%%%%%%%%%%%%%%%%%%%%%%%%%%%%%%%%%%%%%%%%%
\subsection{Stokes approximation}

Within glaciers, the ice behaves as a very slow-moving fluid so that
inertial forces can be neglected and thus
\begin{linenomath}
\begin{gather*}
  \rho(\bmu_t + \bmu\cdot\nabla \bmu) \approx 0,
\end{gather*}
\end{linenomath}
which when substituted into~\eqref{stokes1} yields the linearized Stokes
approximation.  Furthermore, ice flow models typically use a
strain-rate-dependent viscosity to describe ice rheology, for which the
most widely used form (called Glen's flow law) is derived by fitting
empirical measurements of ice deformation to a power
law~\cite{cuffey2010physics}. The standard model for isothermal ice flow
is therefore given by an augmented version of the Stokes equations:
\begin{linenomath}
\begin{alignat}{3}
  \textrm{(incompressibility)} &\qquad& \nabla \cdot \bmu &= 0, 
  \label{eq:incompressible}\\ 
  \textrm{(Stokes equations, stress balance)} && 0 &= -\nabla p + \nabla\cdot
  \tau_{i,j} + \rho \myvec{g}, \label{eq:siastress}\\ 
  \textrm{(Glen's flow law)} &&  \mathcal{D} u_{i,j} &= A \tau^{\alpha-1} 
  \tau_{i,j}, \label{eq:glen} 
\end{alignat}
\end{linenomath}
where the strain rate tensor is
$\mathcal{D} u_{i,j} = \frac{1}{2}\left( \frac{\partial u_i}{\partial
    x_j} + \frac{\partial u_j}{\partial x_i}\right)$ and
$\tau^2 = \frac{1}{2} \tau_{i,j} \tau_{j,i}$ (with the summation
convention assumed for repeated indices). A typical value of the
exponent in Glen's flow law is $\alpha=3$.  Because the Stokes equations
contain no time derivative, the ice velocity $\bmu$, pressure $p$, and
stress field $\tau_{i,j}$ are determined simultaneously from the
gravitational force $\rho \myvec{g}$, ice softness $A$, and boundary
stresses.

%%%%%%%%%%%%%%%%%%%%%%%%%%%%%%%%%%%%%%%%%%%%%%%%%%%%%%%%%%%%%%%%%%
\subsection{Shallow shelf approximation (SSA)}
\label{sec:ssa}

%\begin{figure}[bthp]
%  \centering
  %\includegraphics[width=0.6\textwidth]{figures/icesheetcartoon6}
%  \includegraphics[trim=0cm 5.5cm 0cm 0cm, height=3.5cm, width=9.0cm]{figures/icesheet_cartoon}
  %\includegraphics[width=0.6\textwidth]{figures/alamgirFig1mod}
%  \caption{Geometry of the shallow ice sheet flow problem.\label{fig:cartoon_icesheet_sia}} 
%\end{figure}

\begin{figure}[bthp]
  \centering
  \includegraphics[width=0.5\textwidth]{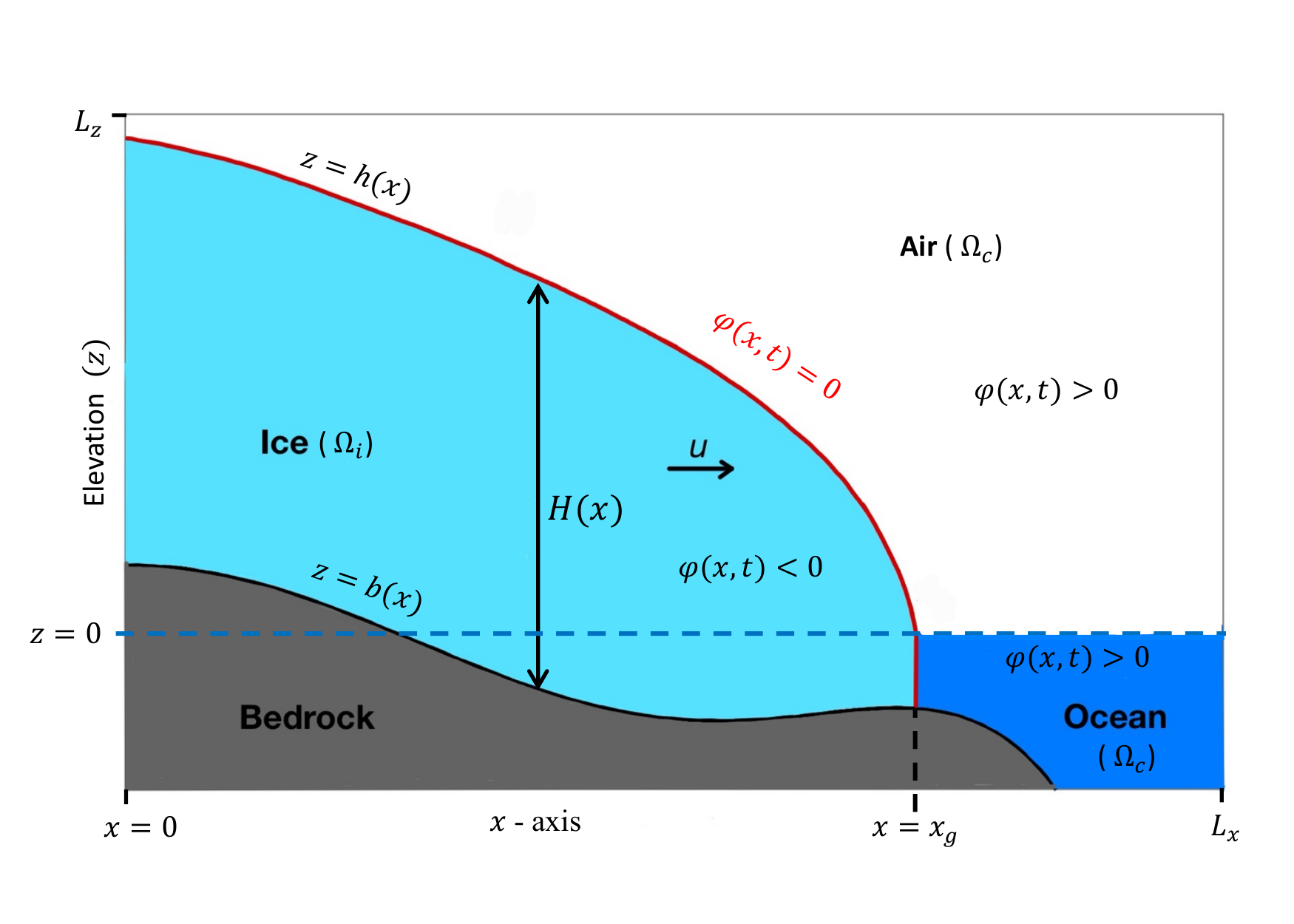}
  \caption{A grounded, marine-terminated ice sheet profile, 
    illustrating how the level set function $\varphi(\bmx,t)$ separates
    the domain into three regions: the ice region ($\Omega_i$ with
    $\varphi<0$), the air/ocean region lying outside the ice ($\Omega_c$
    with $\varphi>0$), and the interface between them ($z=h=H+b$ along
    $\varphi=0$).}  
  \label{fig:cartoon_icesheet}
  % Image taken from Schoof~\cite{schoof2007marine}
\end{figure}

The Stokes equations~\eqref{eq:incompressible}--\eqref{eq:glen} remain
challenging to implement on a routine basis.  The SSA is a
well-established and frequently used approximation that neglects
vertical shear~\cite{macayeal1989large}, which is valid for ice shelves
and ice streams because they are characterized by low basal drag. We
introduce the SSA here for a two-dimensional grounded shallow marine ice
sheet as pictured in Fig.~\ref{fig:cartoon_icesheet}.  Let $x$ denote
the horizontal coordinate in the flow direction with $x=0$ defining the
up-glacier boundary of the ice stream, and let $z$ be the vertical
coordinate measured upward from sea level.  In the horizontal direction,
the ice sheet extends from $x=0$ to the location $x=x_\mathrm{g}$ that
separates grounded ice from the surrounding ocean, which is known as the
grounding line (and terminus position, if a floating ice shelf is
absent).  We then define $H(x,t)$ as the ice thickness at position $x$
and time $t$, and $b(x)$ as the bedrock surface height, so that the
location of the upper ice surface in contact with the atmosphere is at
$z=h(x,t) = H(x,t)+b(x)$ (see Fig.~\ref{fig:cartoon_icesheet}).

% A non-linear friction law that links basal shear stress to basal
% sliding velocity is applied on grounded ice,
% \begin{gather}
%   \tau_{\mathrm{b}} = C |u_{\mathrm{b}}|^{\beta-1}
%   u_{\mathrm{b}}, \label{ssa:frictionlaw} 
% \end{gather}
% where $\tau_\mathrm{b}$ is the basal shear stress, $u_\mathrm{b}$ is
% the basal sliding velocity, $C$ is the friction coefficient, and $\beta$
% is the sliding law exponent. The eq. \eqref{ssa:frictionlaw}) is also
% written as  $\tau_\mathrm{b} = \beta^2 u_\mathrm{b}$ where $\beta^2$
% is known as basal sliding coefficient. 
% \begin{gather*}
%   \left(2A^{-1/\alpha}H|u_x|^{1/\alpha-1}u_x \right)_x - C
%   |u|^{\beta-1} u = \rho g H h_x 
% \end{gather*}

% The momentum conservation for the SSA model originally derived by
% MacAyeal~\cite{macayeal1989large} is as follows, 
The SSA model for a grounded ice sheet occupying the region $0 < x <
x_\mathrm{g}(t)$ is a nonlinear differential equation
\begin{linenomath}
\begin{gather}
  \left(2A^{-1/\alpha} H |u_x|^{1/\alpha - 1} u_x \right)_x -
  C |u|^{\beta-1} u = \rho g H h_x, \label{eq:ssa1} 
\end{gather}
\end{linenomath}
where we note that the spatial domain is time-dependent owing to changes
in the terminus position $x_\mathrm{g}(t)$.  This is a statement of
momentum conservation that is obtained from
Eqs.~\eqref{eq:incompressible}--\eqref{eq:glen}.  The first term in
Eq. \eqref{eq:ssa1} is the vertically-integrated longitudinal
stress. The second term $\tau_\mathrm{b} = C |u|^{\beta-1} u$ represents
the basal shear stress, where $C$ is a friction coefficient (with
$C\equiv 0$ for an ice shelf) and the friction exponent $\beta$ has a
typical value of $\beta=\frac{1}{3}$. The final term on the right hand
side is the driving stress due to gravity,
$\tau_\mathrm{d} = \rho g H h_x$. As a result, the SSA equation
represents a stress balance wherein longitudinal strain rates are
determined by the integrated ice hardness (through the coefficient
$2A^{-1/\alpha}H$), the slipperiness of the bedrock (friction
coefficient $C$ and exponent $\beta$), and the geometry of the ice sheet
(thickness $H$ and surface slope $h_x$).

Because Eq.~\eqref{eq:ssa1} is a 1D nonlinear second-order equation for
$u$, two boundary conditions are required. We also note that effects of
time variation only appear implicitly through changes in the ice
thickness $H(x,t)$. For this reason, we will often write the ice
velocity $u(x,t)$ as $u(x)$, treating the time $t$ as a parameter. On
the left-hand boundary or the up-glacier boundary, the ice is simply
assumed to have a constant velocity:
\begin{linenomath}
\begin{gather}
  u = u_\mathrm{in} \quad \textrm{at} \quad x = 0.\label{eq:ssa4}
\end{gather}
\end{linenomath}
At the terminus position, 
% the ice sheet model is coupled with the ice shelf model. At this point the 
the right-most point where ice begins to float, we impose the flotation
condition
\begin{linenomath}
\begin{gather*}
  h = (1 - \rho/\rho_\mathrm{w}) H \quad \textrm{at} \quad x = x_\mathrm{g}.
\end{gather*}
\end{linenomath}
%At the grounding line $x = x_\mathrm{g}$, the ice just becomes afloat. 
Upon substituting this expression into \eqref{eq:ssa1} and neglecting basal
friction for floating ice (by dropping the term $C
|u|^{\beta-1} u$), we obtain the corresponding right hand boundary
condition for $u$ 
\begin{linenomath}
\begin{gather}
  2A^{-1/\alpha}H|u_x|^{1/\alpha - 1}u_x  = \frac{1}{2} \rho (1 -
  \rho/\rho_\mathrm{w}) g H^2 \quad \textrm{at} \quad x =
  x_\mathrm{g}. \label{eq:ssa3}  
\end{gather}  
\end{linenomath}

The SSA model only determines the horizontal ice velocity $u(x,t)$, but
any ice sheet travelling over a non-uniform bedrock layer will also
induce a vertical ice motion.  We can easily obtain this vertical
velocity of the ice stream by assuming that the bedrock is rigid and
that no
melting occurs at the bottom surface. In that case, we can impose a
kinematic boundary condition that the ice velocity directed
perpendicular to the rigid bedrock must vanish on the boundary itself
($\ddx{b}{t} = 0$).  When combined with the incompressibility condition
\eqref{eq:incompressible}, this yields the vertical velocity
corresponding to any horizontal velocity field $u$ as
\begin{linenomath}
  \begin{gather}
    w(x,z,t) = u(x,t) b_x - (z-b(x)) u_x
    \quad \textrm{for} \quad 0 \leqslant x \leqslant
    x_\mathrm{g}, \label{eq:verticalvelssa} 
  \end{gather}
\end{linenomath}
where the 2D SSA velocity is $\bmu(x,z,t) = \Big(
u(x,t),\, w(x,z,t) \Big)$.

%%%%%%%%%%%%%%%%%%%%%%%%%%%%%%%%%%%%%%%%%%%%%%%%%%%%%%%%%%%%%%%%%%
\subsection{Numerical solution of the SSA}

Taking the ice thickness as a given function $H(x)$, our aim is to solve
the nonlinear differential equation \eqref{eq:ssa1} for the unknown
velocity $u(x)$.  We choose a spatial grid with constant spacing
$\Delta x$ and $N_x$ equally-spaced points $x_i=i\Delta x$, and define
corresponding solution approximations as $u_i\approx u(x_i)$, where
$N_x\Delta x \leqslant x_\mathrm{g} < (N_x+1)\Delta x$.  Note that the
SSA is a quasi-steady 1D approximation used to determine the velocity
components $(u, w)$ in the $x$ and $z$ directions at any time $t$ based
on the current height profile $H(x)$. At discrete times labelled $t^k$, velocity
components are denoted by $u^k$, $w^k$, and ice thickness and surface
height by $H^k$, $h^k$. For simplicity of notation, we will often
suppress the superscript $k$ in the SSA algorithm.

% We divide the interval $[0,x_g]$ into $N_x$ equally-spaced points
% $x_i=i\Delta x$ with $\Delta x=x_g/N_x$ and define approximate
% solution values $u_i\approx u(x_i)$. All derivatives in
% \eqref{eq:ssa1} are approximated by \dots\ [DETAILS HERE]. 

The discrete equations for $u_i$ are solved using a Picard iteration
approach that linearizes the nonlinear system.  Letting $u^{(r)}$ denote
the current velocity iterate, the Picard approach linearizes the
equation by labeling various terms in the left hand side
of~\eqref{eq:ssa1} as follows:
\begin{linenomath}
\begin{gather}
  \left(W^{(r-1)} u^{(r)}_x \right)_x -
  C\left|u^{(r-1)}\right|^{\beta-1} u^{(r)} = \rho g H 
  h_x, \label{eq:picard1}   
\end{gather}
\end{linenomath}
where $W^{(r-1)} = 2 A^{-1/\alpha} H \left|
  u^{(r-1)}_x\right|^{1/\alpha-1}$, and noting that $\rho g H h_x$
remains constant for all iterations. For stability and accuracy reasons,
the coefficient function $W^{(r)}$ is discretized with a compact,
centered difference approximation so that the discrete form of the first
term in \eqref{eq:picard1} is
\begin{linenomath}
\begin{gather*}
  \left(W^{(r-1)} u^{(r)}_x \right)_x = \frac{W_{i+1/2}^{(r-1)}
    (u_{i+1}^{(r)} - u_i^{(r)}) - W_{i-1/2}^{(r-1)} (u_i^{(r)} -
    u_{i-1}^{(r)})}{\Delta x^2}     
\end{gather*}
\end{linenomath}
where
$W_{i+1/2}^{(r-1)} = 2 A^{-1/\alpha} \left(\frac{H_{i+1} +
    H_i}{2}\right) \left| \frac{u_{i+1}^{(r-1)} - u_i^{(r-1)}}{\Delta
    x}\right|^{1/\alpha-1}$. The corresponding discrete form of
\eqref{eq:picard1} is a linear system for $u_i^{(r)}$ that is solved in
each iteration.  In practice, we use a convergence tolerance of
$10^{-10}$ in the max-norm which requires between 5 and 20 Picard
iterations.

To obtain the initial velocity iterate $u^{(0)}$, we assume that the
grounded ice ($0 < x < x_\mathrm{g}$) is held in place by basal
resistance only.  Therefore,
$u^{(0)}(x) = \left( - C^{-1} \rho g H h_x \right)^{1/\beta} $ and the
boundary condition from \eqref{eq:ssa4} is imposed on the left.  The
right end condition \eqref{eq:ssa3} can be written as
\begin{linenomath}
\begin{gather}
  u_x(x_\mathrm{g}) = \frac{u_{N_x + 1} - u_{N_x}}{\Delta x} = A
  \left(\frac{1}{4} \rho (1 - \rho/\rho_\mathrm{w}) g
    H(x_\mathrm{g})\right)^\alpha\,\, \mathrm{where} 
  \,\,N_x\Delta x \leqslant x_\mathrm{g} < (N_x+1)\Delta x.
  \label{eq:picard_bc2}   
\end{gather} 
\end{linenomath}
The discretized form of \eqref{eq:picard1} at $x = x_{N_x}$ contains the
value $u_{N_x+1}$, which is replaced using the relation
\eqref{eq:picard_bc2} and then the ``ghost'' value $W_{N_x+1/2}$ can be
determined using cubic extrapolation.

The vertical velocity is discretized using  centered difference
approximations for $b_x$ and $u_x$ to give
\begin{linenomath}
\begin{gather*}
  w_{i,j} = u_i \frac{b_{i+1} - b_{i-1}}{2 \Delta x} - (z_j-b_i)
  \frac{u_{i+1} - u_{i-1}}{2 \Delta x}, 
  \label{eq:verticalvelssa_discrete}  
\end{gather*} 
\end{linenomath}
with the boundary points handled using one-sided differences. Here, the
vertical coordinate is discretized at points $z_j = j\Delta z$ with
$j=0,1,\dots, N_z$ with $N_z=100$ points in practice.  Because care has
been taken to employ second-order difference approximations for all
spatial derivatives, the SSA algorithm can be expected to converge
with second-order accuracy in space, % $\mathcal{O}(\Delta x^2)$,
which has been confirmed in previous work~\cite{hossain2020modelling}.

\begin{table}[btph]
  \centering
  \begin{tabular}{cccc}\hline
    {\bf Parameter} & {\bf Symbol} & {\bf Value} & {\bf Units} \\\hline
    Ice density & $\rho$  & 900 & $\mathrm{kg/m^3}$ \\
    Water density & $\rho_\mathrm{w}$ & 1000 & $\mathrm{kg/m^3}$ \\
    Ice softness & $A$ & $5.6\times 10^{-17}$ & $\mathrm{1/(Pa^3\,yr)}$ \\
    Gravitational acceleration & $g$ & 9.8 & $\mathrm{m/s^2}$ \\
    Glen's law exponent & $\alpha$ &  3  & -- \\
    Friction exponent   & $\beta$  & 1/3 & -- \\
    Friction coefficient & $C$ & [0.01, 0.015]~\mybar~0.03 & 
    $\mathrm{MPa\,(yr/m)^{1/3}}$ \\
    Velocity at up-glacier boundary & $u_\mathrm{in}$ & 8.0~\mybar~4.5 & $\mathrm{km/yr}$ \\
    Front stress perturbation coefficient & $C_F$  & 1.0~\mybar~[1.5, 3.5] & -- \\
    Domain length & $L_x$  & 10~\mybar~40 & $\mathrm{km}$ \\
    Domain height & $L_z$  & ~~400~\mybar~1000 & $\mathrm{m}$ \\
    % \hline
    Vertical grid dimension& $N_z$ & 100 & -- \\
    Number of ensembles    & $N_e$ & 50  & -- \\
    % Size of state vector   & $N_z$ & --  & -- \\
    % Number of observations & $N_y$ & --  & -- \\
    \hline
  \end{tabular}
  \caption{Values of the physical and numerical parameters.  In the
    ``value'' column, when two numbers are separated by a
    ``\,\mybar\,'', the first number corresponds to Experiments~1 and~2,
    and the second to Experiment~3.}
  \label{tab:params}
\end{table}

%%%%%%%%%%%%%%%%%%%%%%%%%%%%%%%%%%%%%%%%%%%%%%%%%%%%%%%%%%%%%%%%%%
%%%%%%%%%%%%%%%%%%%%%%%%%%%%%%%%%%%%%%%%%%%%%%%%%%%%%%%%%%%%%%%%%%
\section{Level set method}
\label{sec:lsm} 

When the effect of vertical ice accumulation is incorporated into the
ice velocity obtained from the 1D SSA model, the 
combined ice--air and ice--water interface evolves in two
dimensions.  In this section, we describe a numerical approach called
the level set method (LSM) that is capable of accurately and efficiently
tracking this ice interface in 2D.
% The numerical approach describes in this section using level set
% method (LSM) that accurately track ice-air and ice-water interface.
% This model is studied and verified by Hossain et
% al.~\cite{hossain2020modelling}. The numerical model coupling between
% SSA and LSM is named as SSA-LSM model.
We provide only the essential details of the coupled SSA--LSM approach
and direct the interested reader to Hossain et
al.~\cite{hossain2020modelling} for complete details, including
computational studies based on an extensive suite of test cases.

The level set method was originally developed by Osher and
Sethian~\cite{osher1988fronts} for capturing an evolving interface
$\Gamma(t)$ implicitly in terms of a level set function
$\varphi(\bmx,t)$.  The function $\varphi$ is real-valued and
differentiable on a space-time domain $\Omega \times \mathbb{R}_+$,
where the spatial variable is $\bmx \in \Omega$. The interface
$\Gamma(t)\subset\Omega$ is a curve that evolves in time $t$ and is 
represented as the zero isosurface or level set ${\varphi(\bmx,t) = 0}$,
which propagates at a speed directed normal to the interface.

When tracking the interface in our ice sheet model, we employ a common
approach in which the level set function is defined as a signed distance
function
\begin{linenomath}
\begin{gather} 
  \label{eq:signeddistance1}
  \varphi(\bmx, t) = \left\{
    \begin{array}{cl}
      - d(\bmx, \Gamma), & \mathrm{for} \,\,\bmx\in \Omega_i,\\
      \phantom{-} d(\bmx,   \Gamma), & \mathrm{for} \,\,\bmx\in \Omega_c,\\
     0, & \mathrm{for} \,\,\bmx\in \Gamma,
    \end{array}
  \right.
\end{gather}
\end{linenomath}
where $\Omega_i$ represents the region inside the ice body, $\Omega_c$
is the region outside the ice (consisting of either air or water),
$\Gamma$ is the ice--air or ice--water interface, and $\Omega = \Omega_i
\cup \Omega_c \cup \Gamma$ (see Fig.~\ref{fig:cartoon_icesheet}).  The
value of the level set function corresponds to the Euclidean distance
${d(\bmx, \Gamma)}$ between any spatial location $\bmx$ and the
corresponding closest point on the interface $\Gamma$, with the sign
chosen to be negative at points inside the ice region and positive
outside.

% \begin{figure}[bthp]
%   \centering
%   \includegraphics[width=0.5\linewidth]{figures/lsmcartoon4}
%   \caption[]{Basic geometry and definition of the level set
%     function $\varphi(\bmx,t)$ for a generic ice
%     sheet.}\label{fig:cartoonicesheet} 
% \end{figure}

%%%%%%%%%%%%%%%%%%%%%%%%%%%%%%%%%%%%%%%%%%%%%%%%%%%%%%%%%%%%%%%%%%
\subsection{Level set evolution}\label{subsec:lsevo} 
 
In order to derive an equation for the evolution of the level set
function, we first require that any point $\bmx(t) \in \Gamma(t)$ on the
interface being tracked must satisfy $\varphi(\bmx(t), t) = 0$, 
which can be differentiated in time to obtain
\begin{linenomath}
\begin{gather*}
  \ddx{\varphi}{t} + \nabla \varphi(\bmx(t),t) \cdot \bmx'(t) = 0.
\end{gather*}
\end{linenomath}
Let $S$ represent the speed of the interface in the outward-pointing 
normal direction.  Then 
\begin{linenomath}
\begin{gather*}
  \bmx'(t) \cdot \myvec{n} = S \quad \textrm{where} \quad \myvec{n} =
  \nabla \varphi/\|\nabla \varphi\|,  
\end{gather*}
\end{linenomath}
and the evolution equation for the level set function may be re-written
in the form of a Hamilton-Jacobi equation
\begin{linenomath}
\begin{gather}\label{eq:lsee}
  \ddx{\varphi}{t} + S \|\nabla \varphi\| = 0.
\end{gather}
\end{linenomath}

\begin{figure}[bthp]
  \begin{center}
    \includegraphics[height=5.0cm,width=6.0cm]{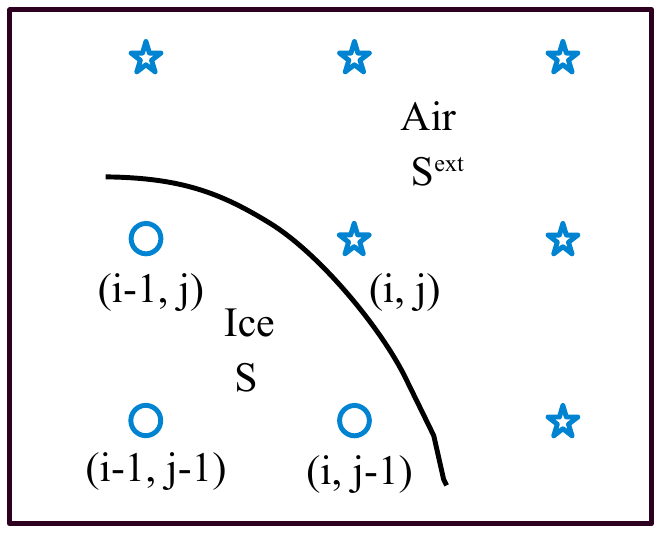}
    \caption[]{\centering The ice--air interface or zero level set of
      $\varphi$ is represented by a solid line, lying in the
      neighbourhood of a discrete point at location $(i,j)$.  The
      interface speed $S$ is known at all points inside the ice, denoted by
      `\textcolor{NavyBlue}{{\Large $\circ$}}' and the speed is 
      extended outside the ice region to obtain $\Sext$ at
      points `\textcolor{NavyBlue}{\large\ding{73}}'.}
    \label{fig:extendedspeed1}
  \end{center}
\end{figure}

The essential remaining component of the LSM is an expression for the
normal speed $S$, which must incorporate the ice velocity field
$\bmu(\bmx,t)$ together with any accumulation or melting along
$\Gamma$~\cite{pralong2004level}. Let $\omega(\bmx,t)$ denote the net
accumulation rate of ice at the upper surface, which is a combination of
snow accumulation and melting (or ablation) that can either be modelled
separately or specified using observational data.  If we assume that
snowfall and melting correspond to vertical variations only, then the
total ice velocity may be written as
$\bmu(\bmx,t) + \omega(\bmx,t) \myvec{\hat{z}}$ where $\myvec{\hat{z}}$
is the unit vector in the vertical direction.  However, this velocity
field is defined only inside the ice region $\Omega_i$, and because the
level set equation is solved throughout the entire computational domain
$\Omega$ we must also determine values of the so-called extended speed
$\Sext$ at points in $\Omega_c$ lying outside the ice region (see
Fig.~\ref{fig:extendedspeed1}).  % A complete description of the
%procedure for computing $\Sext$ at points outside the ice
%region is provided in Hossain et al.~\cite{hossain2020modelling}, based
%on which 
Therefore, we can write the normal speed function as
% \begin{gather}\label{eq:lsspeed}
%   S = (\bmu(\bmx,t) + \omega(\bmx,t) \myvec{\hat{z}}) \cdot \frac{\nabla
%     \varphi}{\|\nabla \varphi\|}, 
% \end{gather}  
\begin{linenomath}
\begin{gather} 
  \label{eq:lsspeed}
  S(\bmx, t) = \left\{
    \begin{array}{ll}
      \Big(\bmu(\bmx,t) + \omega(\bmx,t) \myvec{\hat{z}}\Big) 
      \cdot \myvec{n}, 
      %\frac{\nabla \varphi}{\|\nabla \varphi\|}, 
      & \mathrm{for} \,\,\bmx\in \Omega_i \cup \Gamma,\\ 
      \Sext, & \mathrm{for} \,\,\bmx\in \Omega_c.
    \end{array}
  \right.
\end{gather}
\end{linenomath}

We now complete the LSM algorithm description by illustrating how the
extended velocity field is computed, which is based on the desirable
assumption that $\Sext$ should propagate level sets in such a
way that the signed distance function is
preserved. Following~\cite{zhao1996variational}, $\varphi(\bmx(t),t)$
remains a signed distance function if and only if
\begin{linenomath}
\begin{gather}\label{eq:lsee_ext1}
  \nabla \Sext \cdot \nabla \varphi = 0. 
\end{gather}
\end{linenomath}
Recognizing that interfaces evolving according to \eqref{eq:lsee} may
undergo topological changes, there are a large number of possible cases
that must be considered relating to the configuration of the interface
in the neighbourhood of any discrete point \cite{sethian1999level}.  We
illustrate how to implement \eqref{eq:lsee_ext1} for the special case
pictured in Fig.~\ref{fig:extendedspeed1}, where the point $(i,j)$
lies outside the ice region and $\Sext$ is determined there based
on the three neighbouring point values at $(i-1,j)$, $(i-1,j-1)$ and $(i-1,j-1)$,
all of which lie inside the ice region.  Using a finite difference
discretization of the derivatives in \eqref{eq:lsee_ext1} we can
approximate the extended speed $\Sext$ at position ${(i,j)}$ in terms of
interior values of $\varphi$ and $S$ as
\begin{linenomath}
\begin{gather*}
  \Sext_{i,j} = \frac{S_{i-1,j}
    (\varphi_{i,j}-\varphi_{i-1,j}) + \left(\frac{\Delta x}{\Delta
        z}\right)^2 S_{i,j-1}
    (\varphi_{i,j}-\varphi_{i,j-1})}{(\varphi_{i,j}-\varphi_{i-1,j}) +
    \left(\frac{\Delta x}{\Delta
        z}\right)^2(\varphi_{i,j}-\varphi_{i,j-1})}, 
  \label{eq:lsee_fd}
\end{gather*}
\end{linenomath}
where $\Delta x$ and $\Delta z$ are the horizontal and vertical grid
spacings respectively.  Similar formulas can be derived for other
configurations of the interface and grid to ensure that
Eq.~\eqref{eq:lsee_ext1} is satisfied at all points outside the ice.

%%%%%%%%%%%%%%%%%%%%%%%%%%%%%%%%%%%%%%%%%%%%%%%%%%%%%%%%%%%%%%%%%%
\subsection{Numerical scheme for level set method}\label{sec:Numericalscheme}

% The LSM is a versatile numerical technique that can be implemented in
% concert with a variety of discretizations including finite
% differences, inite elements, moving meshes, etc.  For the sake of
% simplicity, we
We have implemented the level set approach using a fixed, rectangular,
equally-spaced mesh with discrete solution values
$\varphi_{i,j} \approx \varphi(x_i, z_j)$, where the grid spacings
$\Delta x$, $\Delta z$ in each direction may be different.  With similar
approximations of $S_{i,j}$, we approximate the spatial derivatives
$(\varphi_x)_{i,j}$ and $(\varphi_z)_{i,j}$ at position $(x_i, z_j)$ in
Eq.~\eqref{eq:lsee} using the second-order accurate Essentially
Non-Oscillatory (ENO) scheme~\cite{osher2006level}.  For the time
discretization, we employ the Total Variation Diminishing Runge-Kutta
(TVD-RK) scheme of order two, which is a simple explicit method that
determines new values of $\varphi_{i,j}^{k+1}$ at time $t^{k+1}$ based
on known solution values at the previous time $t^k$. The TVD-RK scheme
of second order (also called the modified Euler
method~\citep{osher2006level}) is
\begin{linenomath}
\begin{gather*}
  \varphi_{i,j}^{k+1} = \varphi_{i,j}^{k} + \frac{1}{2} \, \Delta t \,
  S_{i,j}^k \left(\|\nabla\varphi_{i,j}^k\| +
    \|\nabla\widetilde{\varphi}_{i,j}^{k+1}\|\right),
  % \label{eq:lsee_numeric}
\end{gather*}
\end{linenomath}
where
\begin{linenomath}
\begin{gather*}
  \widetilde{\varphi}_{i,j}^{k+1} = \varphi_{i,j}^{k} + \Delta t \, S_{i,j}^k 
  \|\nabla\varphi_{i,j}^k\|
  \quad \text{and} \quad 
  \|\nabla\varphi_{i,j}\| = \sqrt{(\varphi_x)_{i,j}^2 + (\varphi_z)_{i,j}^2}.
\end{gather*}
\end{linenomath}
% the speed in the outward normal direction $S$,
% and the gradient $\nabla \varphi$. 
% We implement the
% Total Variation Diminishing Runge-Kutta (TVD-RK) scheme of second
% order. 
As is usual for explicit schemes of this type, a Courant-Friedrichs-Lewy
(CFL) stability restriction is used to determine the time step $\Delta
t$ in terms of the grid spacings $\Delta x$, $\Delta z$ and the
maximum flow speed. We therefore define
\begin{linenomath}
\begin{gather*}
  \CFL := \frac{\Delta t \, \max_{i,j}|S_{i,j}|}{\min(\Delta x,
    \Delta z)},  
\end{gather*}
\end{linenomath}
and then choose $\Delta t$ small enough so that the condition
$0<\CFL<1$ is always satisfied. 

% In this approach, a higher order non-oscillatory interpolant for
% piecewise smooth functions is used to approximate $\varphi$, and is
% then differentiated piecewise to obtain a corresponding discrete
% approximation for $\nabla \varphi$.  In essence, the ENO approach
% extends first-order accurate upwind differencing to second-order
% spatial accuracy in a way that suppresses oscillations.

The level set is evolved over a time interval corresponding to the next
data assimilation update step (described in the next section).  After
the data assimilation procedure is used to update the ice profiles, the
level set function must be reinitialized according to the signed
distance function. For this purpose, we employ the Fast Marching Method
(FMM)~~\cite{hossain2020modelling} which has proven to be an efficient
algorithm for this reinitialization step.  Indeed, the FMM is capable
of rebuilding $\varphi$ with a computational cost of only
$\mathcal{O}(N\log N)$, where $N=N_x\cdot N_z$ is the total number of
grid points~\cite{adalsteinsson1999fast}.
 
%%%%%%%%%%%%%%%%%%%%%%%%%%%%%%%%%%%%%%%%%%%%%%%%%%%%%%%%%%%%%%%%%%
%%%%%%%%%%%%%%%%%%%%%%%%%%%%%%%%%%%%%%%%%%%%%%%%%%%%%%%%%%%%%%%%%%
\section{Data assimilation}
\label{sec:da}

Data assimilation methods combine observations with a discrete dynamical
model that we write in the general form
\begin{linenomath}
\begin{gather}
  \myvec{z}_{n+1} = \mathcal{M}_n(\myvec{z}_n) + \bm{\zeta}_n,
  \label{eq:damodel1}
\end{gather}
\end{linenomath}
where $\myvec{z}_n \in \mathbb{R}^{N_z}$ is the state vector that
consists of the model solution at each analysis time $t_n$. Note that
the data assimilation procedure is not applied at every time step used
in the underlying discrete model, but rather only at a subset of times
$t_n$ (called analysis times) for which observations are
available. %; so in this section, index $n$ is restricted to this subset.
The model $\mathcal{M}_n$ describes the evolution of the state variables
from time $t_n$ to $t_{n+1}$ and $\bm{\zeta}_n \in \mathbb{R}^{N_z}$
represents a vector of model errors. In this paper, the operator
$\mathcal{M}_n$ corresponds to the process of integrating the discrete
SSA--LSM model up to current analysis time, and the state variables
correspond to the glacier surface heights $H(x_i,t_n)$ and terminus
position $x_\mathrm{g}(t_n)$.
% $\bmu$, $\varphi$, $S$, $H$, $h$ and $x_g$.  
For simplicity, the model is assumed to be a perfect fit so that
$\bm{\zeta}_n \equiv 0$.  The initial state $\myvec{z}_0$ is taken to be
a random vector with given mean $\overline{\myvec{z}}_0 =
E[\myvec{z}_0]$ and covariance $\myvec{P}_0 = E\Big[ (\myvec{z}_0 -
\overline{\myvec{z}}_0)(\myvec{z}_0 - \overline{\myvec{z}}_0)^T \Big]$,
where $E[\cdot]$ denotes the mean or expected value.
 
Suppose in addition that at each analysis time $t_n$ we are provided
with $N_y$ observations denoted $\myvec{y}_n^o$, which are related to
the state vector by the observational equations
\begin{linenomath}
\begin{gather*}
  \myvec{y}_n^o =  \mathcal{H}_n(\myvec{z}_n) + \bm{\eta}_n,
  \label{eq:daobs1}
\end{gather*}
\end{linenomath}
where $\myvec{y}_n^o \in \mathbb{R}^{N_y}$ is called the observation
vector, $\mathcal{H}_n$ is the observation operator that maps the model
space to the observation space, and $\bm{\eta}_n \in \mathbb{R}^{N_y}$
is the measurement noise.  In this paper, we take observations of ice
surface elevation ($h=H+b$) and terminus position ($x_\mathrm{g}$) which
are readily observed by remote sensing methods~\cite{Tedesco_2014}.  We
assume that observations $\myvec{y}_n^o$ and a forecast (or background)
state $\myvec{z}_n^f$ are both provided at time $t_n$. Our aim is then
to compute an analysis state $\myvec{z}_n^a$, which is updated through
the ETKF data assimilation process that we describe next.  Finally, we
use the resulting $\myvec{z}_n^a$ state to obtain the forecast state
$\myvec{z}_{n+1}^f$ by integrating the dynamical model in
Eq.~\eqref{eq:damodel1} to the next analysis time $t_{n+1}$.

%%%%%%%%%%%%%%%%%%%%%%%%%%%%%%%%%%%%%%%%%%%%%%%%%%%%%%%%%%%%%%%%%%
\subsection{Ensemble Kalman Filter}

This paper uses a variant of the Ensemble Kalman Filter (EnKF) which is
a well-known approach for data assimilation based on replacing a single
forecast state with an ensemble of states. The EnKF was first introduced
by Evensen~\cite{evensen1994sequential} and a comprehensive overview of
ensemble filters can be found in~\cite{evensen2003ensemble}. The
ensembles of forecast and analysis states are denoted by
$\Big\{ \myvec{z}_n^{(m)f} \Big\}$ and
$\Big\{ \myvec{z}_n^{(m)a} \Big\}$ for $m=1, 2, \dots, N_e$, where $N_e$
represents the number of ensembles.
% We evolve each ensemble according to the dynamic model
% \eqref{eq:damodel1}, to obtain a forecast ensemble at time $t_{k+1}$:
% \[\myvec{z}_k^{(\alpha)f} = \mathcal{M}(\myvec{z}_{k-1}^{(\alpha)a}).\]
We assume that the number of ensembles is much smaller than the number
of state variables (that is, $N_e \ll N_z$), which is motivated
primarily by limitations on computational resources. The actual forecast
and analysis states are then computed as the average of their respective 
ensembles:
\begin{linenomath}
  \begin{gather}
    \overline{\myvec{z}}_n^f = \frac{1}{N_e} \sum_{m=1}^{N_e} \myvec{z}_n^{(m)f} 
    \quad \text{and} \quad 
    \overline{\myvec{z}}_n^a = \frac{1}{N_e} \sum_{m=1}^{N_e} \myvec{z}_n^{(m)a}. 
    \label{eq:ensemble_xbarkf} 
  \end{gather}
\end{linenomath}
At any time $t_n$ we next define ensemble perturbation matrices based on
these averages as
\begin{linenomath}
  \begin{align}
    \myvec{Z}_n^f &= \left[\left(\myvec{z}_n^{(1)f} -
        \overline{\myvec{z}}_n^f\right), \;
      \left(\myvec{z}_n^{(2)f} - \overline{\myvec{z}}_n^f\right), \; \dots,
      \left(\myvec{z}_n^{(N_e)f} - \overline{\myvec{z}}_n^f\right)\right],
    \label{eq:ensemble_Xkf} \\
    %\text{and} \qquad 
    \myvec{Z}_n^a &= \left[\left(\myvec{z}_n^{(1)a} -
        \overline{\myvec{z}}_n^a\right), \;
      \left(\myvec{z}_n^{(2)a} - \overline{\myvec{z}}_n^a\right), \; \dots,
      \left(\myvec{z}_n^{(N_e)a} - \overline{\myvec{z}}_n^a\right)\right],
    \label{eq:ensemble_Xka} 
  \end{align}
\end{linenomath}
which are both $N_z \times N_e$ matrices whose $m$-th columns are
$\myvec{z}_n^{(m)f} - \overline{\myvec{z}}_n^f$ and
$\myvec{z}_n^{(m)a} - \overline{\myvec{z}}_n^a$ respectively. The
corresponding empirical error covariance matrices can then be written as
\begin{linenomath}
  \begin{align}
    \myvec{P}_n^f &= \frac{1}{N_e-1} \sum_{m=1}^{N_e}
    \left(\myvec{z}_n^{(m)f} - 
      \overline{\myvec{z}}_n^f\right) \cdot
    \left(\myvec{z}_n^{(m)f}-\overline{\myvec{z}}_n^{f}\right)^T
    = \frac{1}{N_e-1}\myvec{Z}_n^f {(\myvec{Z}_n^f)}^T, \\
    \myvec{P}_n^a &= \frac{1}{N_e-1} \sum_{m=1}^{N_e}
    \left(\myvec{z}_n^{(m)a} - \overline{\myvec{z}}_n^a\right) \cdot
    \left(\myvec{z}_n^{(m)a}) - \overline{\myvec{z}}_n^{a}\right)^T
    = \frac{1}{N_e-1} \myvec{Z}_n^a  {(\myvec{Z}_n^a)}^T. 
    \label{eq:ensemble_Pka} 
  \end{align}
\end{linenomath}

The main idea behind the EnKF is to choose initial ensembles at time
$t_n$ whose spread around the average $\overline{\myvec{z}}_{n}$
characterizes the analysis covariance $\myvec{P}_{n}^a$, and then to
propagate each ensemble member using the nonlinear model
$\mathcal{M}_n(\cdot)$ and compute $\myvec{P}_{n+1}^f$ based on the
forecasted ensemble at time $t_{n+1}$~\cite{evensen1994sequential}. 
\begin{figure}[bthp]
  \centering
  \includegraphics[width=0.7\linewidth]{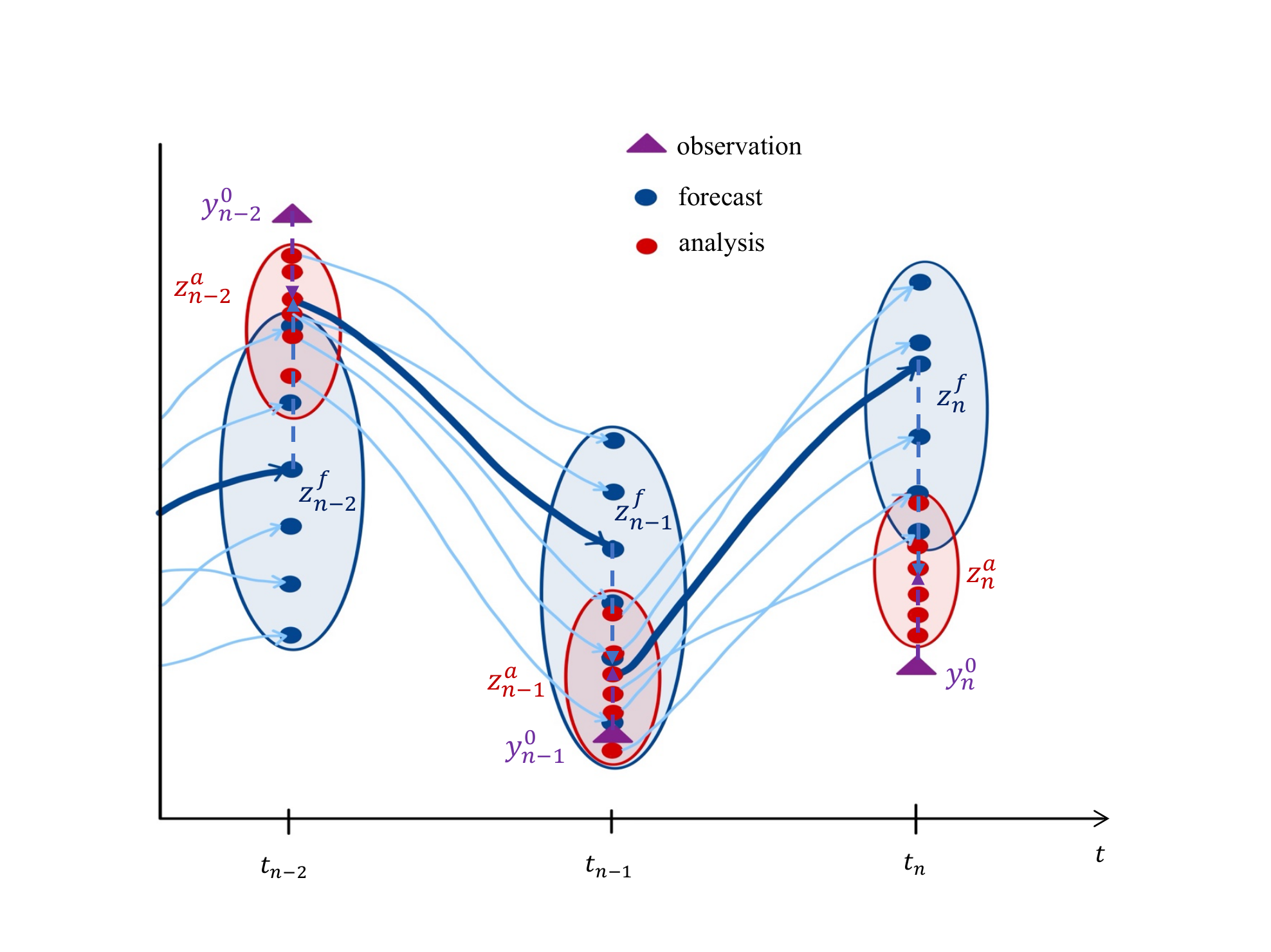}
  \caption{Illustration of the Ensemble Kalman Filter
    (EnKF) sequence. % which is carried out every $n$ time steps. 
    The goal of EnKF is to use available observation (purple triangle)
    to correct the model prediction and get closer to the unknown
    truth. Blue and red ellipses represent the uncertainty in forecast
    and assimilated states, respectively. Thin blue lines depict the
    evolution of ensembles, while the thick blue line traces the
    average or mean state.} 
  \label{fig:data_cartoon_EnKF}
\end{figure}
This EnKF procedure is illustrated graphically in
Fig.~\ref{fig:data_cartoon_EnKF},
%Making use of this notation, 
and the process can be summarized compactly as follows: the EnKF
forecast ensembles are determined by applying the model to the previous
analysis states
\begin{linenomath}
  \begin{gather*}
    \myvec{z}_{n+1}^{(m)f} = \mathcal{M}_{n}(\myvec{z}_{n}^{(m)a}), 
    \label{eq:ensemble_xmkf}
  \end{gather*}
\end{linenomath}
and then the data assimilation step can be written in matrix form as
\begin{linenomath}
  \begin{align*}
    \myvec{K}_{n+1} &= \myvec{P}_{n+1}^f \myvec{H}_{n+1}^T \Big[\myvec{H}_{n+1}
    \myvec{P}_{n+1}^f \myvec{H}_{n+1}^T + \myvec{R}_{n+1} \Big]^{-1},\\ 
    \myvec{z}_{n+1}^{(m)a} &= \myvec{z}_{n+1}^{(m)f} +
    \myvec{K}_{n+1}\left(\myvec{y}_{n+1}^o -
      \mathcal{H}_{n+1}(\myvec{z}_{n+1}^{(m)f})\right). 
  \end{align*}
\end{linenomath}
Here, $\myvec{K}_n$ (with time index shifted by 1 for simplicity) is
constructed in terms of the observational error covariance matrix
$\myvec{R}_n = E[\bm{\eta}_n \bm{\eta}_n^T]$
%\begin{linenomath}
 % \begin{gather}
 %   \myvec{R}_n = [\bm{\eta}_n \bm{\eta}_n^T]    
 % \end{gather}
%\end{linenomath}
and the matrix products
\begin{linenomath}
  \begin{align*}
    \myvec{P}_n^f\myvec{H}_n^T &= \frac{1}{N_e-1} \sum_{m=1}^{N_e}
    \left(\myvec{z}_n^{(m)f} - \overline{\myvec{z}}_n^f\right)
    \left(\mathcal{H}_n(\myvec{z}_n^{(m)f}) - \frac{1}{N_e}
      \sum_{m'=1}^{N_e} \mathcal{H}_n
      \left(\myvec{z}_n^{(m')f}\right)\right)^T, \\
    \myvec{H}_n\myvec{P}_n^f\myvec{H}_n^T &= \frac{1}{N_e-1}
    \sum_{m=1}^{N_e} \left(\mathcal{H}_n(\myvec{z}_n^{(m)f}) -
      \frac{1}{N_e} \sum_{m'=1}^{N_e} \mathcal{H}_n
      \left(\myvec{z}_n^{(m')f}\right)\right)  
    \left(\mathcal{H}_n(\myvec{z}_n^{(m)f}) - \frac{1}{N_e}
      \sum_{m'=1}^{N_e} \mathcal{H}_n
      \left(\myvec{z}_n^{(m')f}\right)\right)^T.   
  \end{align*}
\end{linenomath}
The error covariances $\myvec{P}_n^f$ are propagated implicitly in time
through the ensemble forecasts, which is suitable for large-scale
problems and avoids any need to linearize the model equations.  In
contrast, for a single-forecast filter (without ensembles) the
covariances would be propagated explicitly through the use of linearized
versions of the model and observation operators.

%%%%%%%%%%%%%%%%%%%%%%%%%%%%%%%%%%%%%%%%%%%%%%%%%%%%%%%%%%%%%%%%%%
\subsection{ETKF formulation}

Among the many variants of the EnKF
approach~\cite{evensen2003ensemble}, we apply a
particular deterministic variant known as the Ensemble Transform Kalman
Filter (ETKF) that was originally introduced by Bishop et
al.~\cite{bishop2001adaptive}. The key to the ETKF method is defining
the analysis covariance matrix $\myvec{P}^a$ as a linear transformation
of the forecast ensemble perturbation $\myvec{Z}^f$.  To this end, we
first  write the forecast ensembles projected onto the observation space
in the form of an $N_y \times N_e$ matrix
\begin{linenomath}
  \begin{gather*}
    %\bm{\Upsilon}_n^{f} = 
    \left[\myvec{y}_n^{(1)f}, \; \myvec{y}_n^{(2)f},
      \; \dots , \;
      \myvec{y}_n^{(N_e)f}\right] =
    \left[\mathcal{H}_n(\myvec{z}_n^{(1)f}), \;
      \mathcal{H}_n(\myvec{z}_n^{(2)f}), \; \dots, \;
      \mathcal{H}_n(\myvec{z}_n^{(N_e)f})\right],
  \end{gather*}
\end{linenomath}
where we recall that $\mathcal{H}$ is the observation operator from 
\eqref{eq:daobs1}.  These forecast ensembles have mean  
\begin{linenomath}
  \begin{gather}
    \overline{\myvec{y}}_n^f = \frac{1}{N_e} \sum_{m=1}^{N_e}
    \myvec{y}_n^{(m)f}
    \label{eq:ensemble_ybarkf}  
  \end{gather}
\end{linenomath}
and associated associated perturbation matrix
\begin{linenomath}
  \begin{gather}
    \myvec{Y}_n^f = \left[\left(\myvec{y}_n^{(1)f} -
        \overline{\myvec{y}}_n^f\right), \; \left(\myvec{y}_n^{(2)f} -
        \overline{\myvec{y}}_n^f\right), \; \dots,
      \left(\myvec{y}_n^{(N_e)f} -
        \overline{\myvec{y}}_n^f\right)\right] .
    \label{eq:ensemble_Ykf} 
  \end{gather} 
\end{linenomath}

Next, let $\myvec{q}$ be a vector in $\mathbb{R}^{N_e}$ so that
$\myvec{Z}_n^f\myvec{q}$ belongs to the space $\mathbb{R}^{N_z}$ spanned
by the forecast ensemble perturbation matrix $\myvec{Z}_n^f$, and let
$\myvec{z}_n = \overline{\myvec{z}}_n^f + \myvec{Z}_n^f\myvec{q}$ be the
corresponding model state vector. If $\myvec{q}$ is taken to be a
Gaussian random vector with mean $\myvec{0}$ and covariance $(N_e -
1)^{-1} \myvec{I}$, then $\myvec{z}_n = \overline{\myvec{z}}_n^f +
\myvec{Z}_n^f\myvec{q}$ is also Gaussian with mean
$\overline{\myvec{z}}_n^f$ and covariance $ \myvec{P}_n^f =
\frac{1}{N_e-1}\myvec{Z}_n^f (\myvec{Z}_n^f)^T$.  Following
\cite{houtekamer2001sequential, hunt2007efficient} we make the linear
approximation
\begin{linenomath}
  \begin{gather*}
    \mathcal{H} \Big( \overline{\myvec{z}}_n^f + \myvec{Z}_n^f\myvec{q} \Big)
    \approx \overline{\myvec{y}}_n^f + \myvec{Y}_n^f\myvec{q}
  \end{gather*}
\end{linenomath}
in order to construct a cost function to use in finding an optimal choice
for $\myvec{q}$. 
% [[~What's the point of this? I'm missing something here~]]
The ETKF analysis step then corresponds to minimizing the following cost
function~\cite{hunt2007efficient}
\begin{linenomath}
  \begin{gather*}
    \mathcal{J}(\myvec{q}) = \frac{(N_e - 1)}{2}\myvec{q}^T\myvec{q} +
    \frac{1}{2}\left( \myvec{y}^o - \overline{\myvec{y}}^f -
      \myvec{Y}^f\myvec{q}\right)^T 
    \myvec{R}^{-1}\left( \myvec{y}^o - \overline{\myvec{y}}^f
      -\myvec{Y}^f\myvec{q}\right)    
  \end{gather*} 
\end{linenomath}
(dropping the time index $n$ for simplicity)
% This cost function is in the form of the standard Kalman filter cost
% function, using the background mean $\myvec{0}$ and background
% covariance $(N_e - 1)^{-1} I$.
to obtain the minimizer
\begin{linenomath}
  \begin{align}
    \myvec{q}^a = \arg \min_{\myvec{q}} \, \mathcal{J}(\myvec{q})
    = \widetilde{\myvec{P}}^a (\myvec{Y}^f)^T \myvec{R}^{-1}\left(\myvec{y}^o -
      \overline{\myvec{y}}^f \right), 
    \label{eq:ensemble_wa} 
  \end{align}
\end{linenomath}
where
\begin{linenomath}
  \begin{gather}
    \widetilde{\myvec{P}}^a = \left((N_e - 1) \myvec{I} + (\myvec{Y}^f)^T
      \myvec{R}^{-1} \myvec{Y}^f \right)^{-1}
    \label{eq:ensemble_tildePa}  
  \end{gather}
\end{linenomath}
is the inverse of the Hessian of $\mathcal{J}$ at the minimizer.  In the
model state space, the corresponding analysis mean and covariance are
given by
\begin{linenomath}
  \begin{align}
    \overline{\myvec{z}}^a &= \overline{\myvec{z}}^f +
    \myvec{Z}^f\myvec{q}^a,
    \label{eq:ensemble_xa} \\
    %\text{and} \quad 
    \myvec{P}^a &=\myvec{Z}^f \widetilde{\myvec{P}}^a
    {(\myvec{Z}^f)}^T.
    \label{eq:ensemble_Pa} 
  \end{align}  
\end{linenomath}

To initialize the ensemble forecast that will produce the background
state for the next analysis step, we have to choose an analysis ensemble
whose sample mean and covariance are equal to $\overline{\myvec{z}}^a$
and $\myvec{P}^a$ respectively. We therefore construct a matrix
$\myvec{Z}^a$ so that the sum of its columns is zero and
Eq.~\eqref{eq:ensemble_Pka} holds. To this end, the ensemble mean is
updated using the analysis equation \eqref{eq:ensemble_xa},
whereas the ensemble perturbations are updated using the simple
linear transformation~\cite{bishop2001adaptive} 
% We choose the original formulation of the ETKF
% \cite{bishop2001adaptive}  by  the  ensemble transform matrix
% $\myvec{Q}^a$,  
\begin{linenomath}
  \begin{gather*}
    \myvec{Z}^a = \myvec{Z}^f \myvec{Q}^a,
    \label{eq:ensemble_Xa} 
  \end{gather*} 
\end{linenomath}
where the ensemble transform matrix is
% $\myvec{Q}^a$~\cite{bishop2001adaptive}, 
\begin{linenomath}
  \begin{gather*}
    \myvec{Q}^a = \left((N_e - 1)\widetilde{\myvec{P}}^a\right)^{1/2}.
    \label{eq:ensemble_Wa}
  \end{gather*}
\end{linenomath}
% and by the 1/2 power of a symmetric matrix we mean its symmetric square root. 
It is then straightforward to show that using 
\begin{linenomath}
  \begin{gather*}
    \widetilde{\myvec{P}}^a = (N_e - 1)^{-1}\myvec{Q}^a (\myvec{Q}^a)^T 
  \end{gather*}
\end{linenomath}
in Eq.~\eqref{eq:ensemble_Pka} satisfies~\eqref{eq:ensemble_Pa}, 
%To verify that the sum of the columns of $\myvec{z}^a$ are zero, we can
%check the equivalent expression $\myvec{z}^a \myvec{v} = 0$ where
%$\myvec{v} = (1, 1, \dots , 1)^T$ is a column vector of $N_z$ ones. We
%know from \eqref{eq:ensemble_Ykf} that the sum of columns of
%$\myvec{Y}^f$ is zero, giving 
%\[(\widetilde{\myvec{P}^a})^{-1}\myvec{v} = \left[\left((N_e - 1)
%    \myvec{I} + ({\myvec{Y}^f})^T \myvec{R}^{-1} \myvec{Y}^f
%  \right)^{-1}\right]\myvec{v} = (N_e - 1)\myvec{v}.\] Therefore,
% $\myvec{v}$ is an eigenvector of $\widetilde{\myvec{P}^a}$ with
% eigenvalue $(N_e - 1)^{-1}$. Then by Eq.~\eqref{eq:ensemble_Wa},
% $\myvec{v}$ is also an eigenvector of $\myvec{Q}^a$ with eigenvalue
% 1. Since the sum of the columns of $\myvec{z}^f$ is zero (see
% Eq.~\eqref{eq:ensemble_Xkf}) and $\myvec{Q}^a \myvec{v} = \myvec{v}$
% imply \[\myvec{z}^a\myvec{v} = \myvec{z}^f\myvec{Q}^a\myvec{v} =
% \myvec{z}^f\myvec{v} = 0.\] This transformation of the forecast
% ensemble perturbation matrix $\myvec{Z}^f$ to obtain the analysis
% ensemble perturbation matrix can be generalized by introducing an
% arbitrary orthogonal matrix $\myvec{U}$ into the right hand side of
% Eq.~\eqref{eq:ensemble_Xa} (see \cite{sakov2008implications} for a
% detailed derivation).
after which the analysis
ensemble perturbation matrix can be obtained using
\begin{linenomath}
  \begin{gather}
    \myvec{Z}^a = \myvec{Z}^f \left( (N_e-1)
      \widetilde{\myvec{P}}^a\right)^{1/2} 
    \label{eq:ensemble_Xa_2}  
  \end{gather}  
\end{linenomath}
% where $\myvec{I}$ is the identity matrix that preserves the ensemble mean.
% For the sake of simplicity, we set $\myvec{U}$ equal to the identity
% matrix.
Finally, the analysis ensemble is formed by adding $\overline{\myvec{z}}^a$
to each column of $\myvec{Z}^a$ using 
\begin{linenomath}
  \begin{gather}
    \myvec{z}^{(m)a} = \overline{\myvec{z}}^a +
    \myvec{Z}^{(m)a} , 
    \label{eq:ensemble_xma} 
  \end{gather}  
\end{linenomath}
for $m = 1, 2, \dots, N_e$, after which we may then proceed to the next
forecast step by evolving the nonlinear model \eqref{eq:damodel1}.

%%%%%%%%%%%%%%%%%%%%%%%%%%%%%%%%%%%%%%%%%%%%%%%%%%%%%%%%%%%%%%%%%%
%%%%%%%%%%%%%%%%%%%%%%%%%%%%%%%%%%%%%%%%%%%%%%%%%%%%%%%%%%%%%%%%%%
\section{The SSA--LSM--ETKF algorithm}
\label{sec:da_algorithm}

We now present the SSA--LSM--ETKF procedure described in the previous
three sections in algorithmic form, which combines the SSA--LSM
algorithm for evolving the ice sheet velocity and interface along with
the ETKF algorithm for assimilating observations.  Assuming that
measurements of surface elevation and terminus position are available at a
discrete set of analysis times, we begin by defining the level set
function for each initial ensemble. Starting from the state vector
$\myvec{z}_n$ at time $t_n$ (which contains the terminus position and
ice thicknesses along the glacier flow line), the forecast step proceeds
by integrating the nonlinear model \eqref{eq:damodel1} for each ensemble
over a single time step.
% where we recall that the operator $\mathcal{M}(\cdot)$ evolves the state
% vector using the SSA--LSM model.
% This coupling algorithm between SSA and LSM is the Forecast step. 
At each analysis time, the data assimilation phase applies the ETKF
scheme to generate an analysis state $\myvec{z}^a$, after which the
level set function is reinitialized for each ensemble.  This
forecast--assimilation cycle is expanded in detail below:
\begin{enumerate}[1.]
\item \emph{Initialization step:}  
  % Initialize all $N_e$ ensembles
  % according to Section~\ref{sec:Initial_Ensembles}.
  Build a level set function $\varphi$ for each ensemble using the
  signed distance function in Eq.~\eqref{eq:signeddistance1}, which
  requires computing the shortest distance between each discrete grid
  point $\myvec{x}_{i,j}$ and the ice interface $\Gamma$.
  % of each initial ensemble.  Create a level set function for each
  % ensemble using the signed distance function defined in
  % Eq.~\eqref{eq:signeddistance1} [[ TO WHAT?? discrete grid points
  % $\myvec{x}_{i,j}$ and $\Gamma$ from what? ]].
  To ensure that $\varphi$ is smooth, a finer spatial grid with 50
  regularly-spaced points is used in the interior. To handle the
  possibility that $\Gamma$ may be non-analytic, cubic spline
  interpolation is used to generate discrete values of $\Gamma$ on finer
  grids. 
  % In order to create smooth level set functions, we created a
  % discrete ice interface $\Gamma$ function by taking 50 points inside
  % a grid and used cubic spline interpolation. Then compute the
  % distance from grid points to $\Gamma$.
  
  % To evolve level set function $\varphi$ for each analysis ensemble
  % $\myvec{z}^{(\alpha)a}$ proceed the forecast step to determine
  % forecast state $\myvec{z}^{(\alpha)f}$ using the level set method to
  % the next time where observations are available.
  
\item \emph{Forecast step:}
  % The 2D ice velocity field ${\bmu=(u, w)}$ comes from SSA, see
  % Section~\ref{sec:ssa}.
  For each ensemble, the ice velocities $u$, $w$, speed function $S$,
  and level set function $\varphi$ are computed on a 2D grid of points,
  $\myvec{x}_{i,j}$.  The ice thickness $H$ and upper surface height $h$
  are computed at points $x_i$.
  % Continue the following steps 2.1-2.4 until
  % time $t_{k+1}$ where observations are
  % available from time $t_k$ for each ensemble.
  \begin{enumerate}[2.1] 
  \item Compute velocities $u^k$ and $w^k$: At each time $t^k$, the ice
    thickness $H^k$ and the height of the upper surface $h^k$ are
    known. Use these along with the horizontal derivative of $h^k$
    (determined using a centered difference approximation) in
    Eq.~\eqref{eq:picard1} to determine $u^k$. Then use centered
    approximations of derivatives in Eq.~\eqref{eq:verticalvelssa} to
    determine $w^k$.

  \item Compute normal speeds $S^k$: Use the ice velocities $(u^k,w^k)$
    and the given surface mass balance function $\omega$ to 
    determine $S^k$ at all grid points inside the ice sheet using
    Eq.~\eqref{eq:lsspeed}. At exterior points, values of $S^k$ are
    determined using the procedure described in 
    Section~\ref{subsec:lsevo}.

  \item Compute $\varphi^{k+1}$: The level set function is evolved to
    the next time step using Eq.~\eqref{eq:lsee} and the procedure
    described in Section~\ref{sec:Numericalscheme}.

  \item Extract $H^{k+1}$: The zero level set is determined with subgrid
    scale precision as the zero contour line of $\varphi^{k+1}$, which
    is then used to compute $H^{k+1}$ and $h^{k+1}$.  The terminus
    position $x_\mathrm{g}^k$ is identified as the point where the zero level set
    intersects the bottom boundary of the domain.
    % Although we note that our ice velocity solver does not use any
    % subgrid method; for example, the exact grounding line position is
    % not used in the SSA computation.

  \item Inner time-stepping loop: Set $t^{k+1} = t^k+\Delta t$,
    increment $k$ and return to Step 2.1, until the next analysis time
    $t_{n+1}$.
  \end{enumerate}

\item \emph{Data assimilation step:} At each analysis time $t_{n}$ when
  observations are available, the ETKF procedure combines forecast
  ensemble states $\myvec{z}^{(m)f}$ (consisting of ice thickness and
  terminus position) with observations $\myvec{y}^o$ to obtain analysis
  states $\myvec{z}^{(m)a}$.  This ETKF scheme is performed according to
  the following steps:
  \begin{enumerate}[3.1]
  \item Use forecast ensembles of ice thickness and terminus
    position to determine $\overline{\myvec{z}}^f$ and $\myvec{Z}^f$
    using Eqs.~\eqref{eq:ensemble_xbarkf} and \eqref{eq:ensemble_Xkf}.
  \item Compute the corresponding forecast ensembles projected onto the
    observation space, $\overline{\myvec{y}}^f$ and $ \myvec{Y}^f$,
    using \eqref{eq:ensemble_ybarkf} and \eqref{eq:ensemble_Ykf}.
    % \item Choose observation $\myvec{y}^o$.  
  \item %As we know that the terminus is highly sensitive and
    % could undergo rapid change. For one dimensional ice profile,
    % state and observation of  the terminus is just a point.  
    % First, we proceed with the data assimilation steps 3.4-3.7 for
    % grounding lines which is significant for ice geometry, however,
    % computationally less expensive compare to to the ice thickness
    % state vector. 
    The analysis procedure is divided into two stages, the first of
    which updates the analysis state only for the single element
    corresponding to the terminus position, while holding the ice
    thickness constant:
    \begin{enumerate}[a)]
    \item Compute $\myvec{q}^a$ and $\widetilde{\myvec{P}}^a$ from
      Eqs.~\eqref{eq:ensemble_wa} and \eqref{eq:ensemble_tildePa}.
      % $\eqref{eq:ensemble_tildePa} : \qquad \widetilde{\myvec{P}}^a =
      % \left((N_e-1) \myvec{I} + ({\myvec{y}^f})^T \myvec{R}^{-1}
      %   \myvec{y}^f \right)^{-1}$\\
      % $\eqref{eq:ensemble_wa} : \qquad \myvec{q}^a =
      % \widetilde{\myvec{P}}^a
      % ({\myvec{y}^f})^T \myvec{R}^{-1}\left(\myvec{y}^o -
      %   \overline{\myvec{y}}^f \right)$
    \item Compute $\overline{\myvec{z}}^a$ using $\myvec{q}^a$ and
      forecasts $\overline{\myvec{z}}^f$, $\myvec{Z}^f$ in
      \eqref{eq:ensemble_xa}.
      % $\eqref{eq:ensemble_xa} : \qquad \overline{\myvec{z}}^a =
      % \overline{\myvec{z}}^f + \myvec{z}^f\myvec{q}^a$
    \item Compute $\myvec{Z}^a$ using $\myvec{Z}^f$ and
      $\widetilde{\myvec{P}}$ in \eqref{eq:ensemble_Xa_2}. 
      % , where we set the orthogonal matrix $\myvec{U}=\myvec{I}$.
      % $\eqref{eq:ensemble_Xa_2} : \qquad \myvec{z}^a = \myvec{z}^f
      % \left( (N_e-1)\widetilde{\myvec{P}}^a\right)^{1/2}\myvec{U}$\\
    \item Compute the analysis ensembles by adding
      $\overline{\myvec{z}}^a$ to each of the columns of $\myvec{Z}^a$
      using \eqref{eq:ensemble_xma}.
      % $\eqref{eq:ensemble_xma} : \qquad \myvec{z}^{(\alpha)a} =
      % \overline{\myvec{z}}^a + \myvec{z}^{(\alpha)a}$, where $\alpha =
      % 1, 2, \dots, N_e$. 
    \end{enumerate}

  \item Update the ice thickness to account for calving at the glacier
    front in two cases:
    
    \begin{enumerate}[a)]
    \item If the analysis terminus point is advanced beyond the
      forecasted ice thickness profile, then extend the ice thickness
      profile using linear interpolation.
    \item If the analysis terminus point lies inside the forecasted ice
      profile, then adjust the ice thickness profile by removing all ice
      downstream of the new terminus position.
    \end{enumerate}

  \item In stage two of the analysis procedure, repeat Steps 3.3a--d for
    the elements corresponding to ice thicknesses, while holding the
    terminus position constant.
  \end{enumerate}
  
\item \emph{Reinitialization step:}
  % \begin{enumerate}[4.1]
  % \item 
  The analysis ensemble $\myvec{z}^{(m)a}$ defines the ice thickness
  profile and terminus position for each ensemble. The FMM algorithm
  from Section~3.2 and~\cite{hossain2020modelling} is used to rebuild
  the level set function $\varphi$ for each analysis ensemble
  $\myvec{z}^{(m)a}$.%, $m = 1, 2, \dots, N_e$.
  % \item To evolve level set function $\varphi$ for each analysis ensemble
  %   $\myvec{z}^{(\alpha)a}$ proceed the forecast step to determine forecast
  %   state  $\myvec{z}^{(\alpha)f}$ using the level set method to the next time
  %   where observations are available.  
  % \end{enumerate}

\item \emph{Outer loop:} Return to Step~2 to continue the time
  integration. 
\end{enumerate}

%%%%%%%%%%%%%%%%%%%%%%%%%%%%%%%%%%%%%%%%%%%%%%%%%%%%%%%%%%%%%%%%%%
%%%%%%%%%%%%%%%%%%%%%%%%%%%%%%%%%%%%%%%%%%%%%%%%%%%%%%%%%%%%%%%%%%
\section{Numerical results}
\label{sec:dataassimilation_results}

We now apply the SSA--LSM--ETKF algorithm just described to three
experiments of increasing complexity: 
\begin{itemize}
\item {\normalfont\itshape Experiment 1:} An idealized marine outlet
  glacier undergoing inter-annual advances and retreats in response to a
  given melt rate that is a simple periodic function of time.  This
  first test is performed without any data assimilation to examine the
  SSA--LSM component of the model only.
\item {\normalfont\itshape Experiment 2:} A repeat of the first
  experiment with a more realistic pseudo-random time series for the
  melt rate forcing. Data assimilation is incorporated using
  synthetic ensembles determined by perturbing the background state.
\item {\normalfont\itshape Experiment 3:} A more realistic test of the
  data assimilation algorithm that simulates terminus change at Helheim
  Glacier, one of Greenland's largest outlet glaciers. The data
  assimilation step incorporates actual measurements taken over the
  period 2001--2006. 
\end{itemize}
Each experiment is described separately in the following three sections.

%%%%%%%%%%%%%%%%%%%%%%%%%%%%%%%%%%%%%%%%%%%%%%%%%%%%%%%%%%%%%%%%%%
\subsection{Experiment 1: Idealized seasonal ice dynamics, without data
  assimilation} 
\label{sec:seasonal_noDA} 

Here we consider the SSA model coupled with the LSM (as described in
Section~\ref{sec:da_algorithm}) to track a simple grounded marine ice
sheet profile as it advances and retreats in response to a synthetic
seasonal melting variation.  We ignore any effects of ice melange or
tidal flexure on the ice dynamics, and assume that the glacier
terminates at a calving front with grounding line position
$x = x_\mathrm{g}$ so that there is no floating portion of the ice
shelf.

\subsubsection{Experimental design}

\begin{figure}[bthp]
  \begin{center}
    \includegraphics[width=0.6\textwidth]{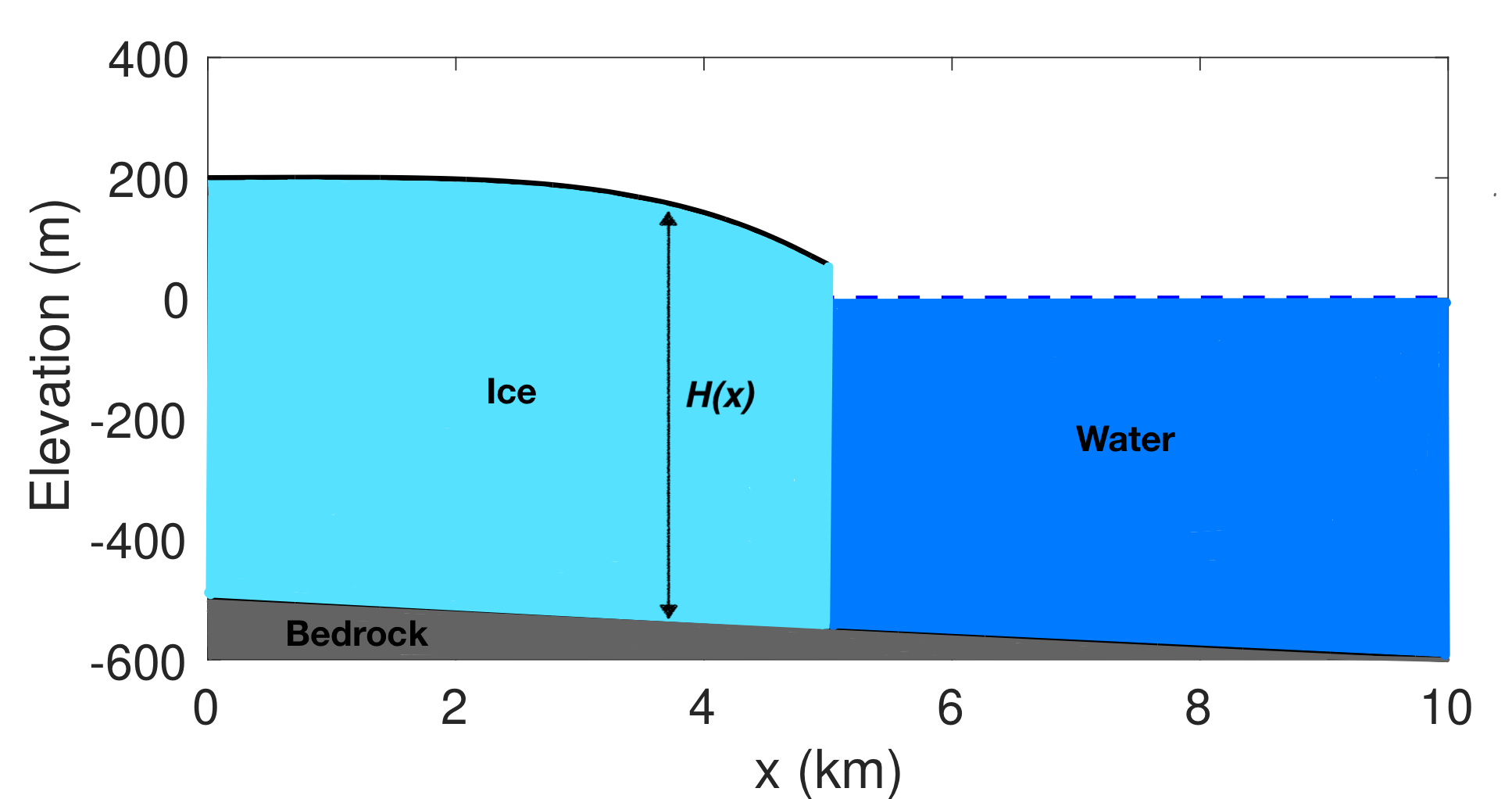}
    \caption{Initial profile of the idealized marine
      outlet glacier in Experiment~1. 
      % Setup of the experiment. The glacier is grounded on a solid
      % bedrock with a slightly downward linear slope. %Grounded ice and
      % water are distinguished with two different colors. 
    }
    \label{fig:initial_seasonal}
  \end{center}
\end{figure}

Following an example from Krug et al.~\cite{krug2015modelling}, the ice
sheet is given an initial profile with surface height
\begin{linenomath}
  \begin{gather} 
    h(x) = h_0 \left(1 - (x/x_\mathrm{g})^4 \right) - (1 - \rho/\rho_\mathrm{w})
    b(x) \quad \text{for}\;0 \leqslant x \leqslant x_\mathrm{g},  
    \label{eq:surfaceseasonal}
  \end{gather}
\end{linenomath}
which lies on top of bedrock having a constant 1\% downward slope described
by the function $b(x) = -500 - 0.01 x$ (measured in m).  The initial
terminus position is at $x_\mathrm{g} = 5000$~m (where the flotation
condition~\eqref{eq:ssa3} is satisfied) and the glacier surface is fixed
on the left boundary at a height $h(0)=200\,\mathrm{m}$ above sea level
throughout.  This initial ice sheet geometry is shown in 
Fig.~\ref{fig:initial_seasonal}.  The other constant parameter values
are ice density $\rho = 900\,\mathrm{kg/m^3}$, water density $\rho_\mathrm{w} =
1000\,\mathrm{kg/m^3}$ and $h_0 = 150\,\mathrm{m}$.  The basal friction
coefficient $C(x)$ from Eq.~\eqref{eq:ssa1} is specified as a linear
function that starts at $1.5 \times 10^{-2}
\,\mathrm{MPa\,m^{-1/3}\,yr^{1/3}}$ at the up-glacier (left) boundary
where $x = 0$~m, and decreases in the flow direction to $1.0 \times
10^{-2} \,\mathrm{MPa\,m^{-1/3}\,yr^{1/3}}$ at $x = 10000\,\mathrm{m}$.
% The total depth-integrated flux through the inlet boundary is kept
% constant ($u_\mathrm{in} = 8.0 \mathrm{km/yr}$).  
The ice softness parameter $A$ is set to $5.6\times10^{-17}\,
\mathrm{Pa^{-3}\,yr^{-1}}$, which is typical of glacial ice rheology at 
a temperature of $-5^{\circ}$C.
% Rignot et al.~\cite{rignot2010rapid} measured summer melt rates between
% 0.6 to $3.8\,\mathrm{m}\,\mathrm{day}^{-1}$ at the face of four calving
% glaciers in West Greenland. Cook et al.~\cite{cook2014modelling} tested
% different maximal melt rate (MMR) from 2.7 to
% $13\,\mathrm{m}\,\mathrm{day}^{-1}$ during the 5-month summer period
% and a constant MMR $0.41\,\mathrm{m}\,\mathrm{day}^{-1}$ or
% $150\,\mathrm{m/yr}$ in winter.  

\begin{figure}[bthp]
  \begin{center}
    %\includegraphics[trim=5cm 5cm 5cm 5cm,width=0.7\textwidth]{figures/meltRateYearly_no_da_2.pdf}\\
    %\includegraphics[trim=5cm 4cm 5cm 6.7cm,width=0.7\textwidth]{figures/GLp-ssa-da2-Seasonal2_2.pdf}
    %trim=0cm 5.5cm 0cm 0cm, height=3.5cm, width=9.0cm
    \includegraphics[width=0.65\textwidth]{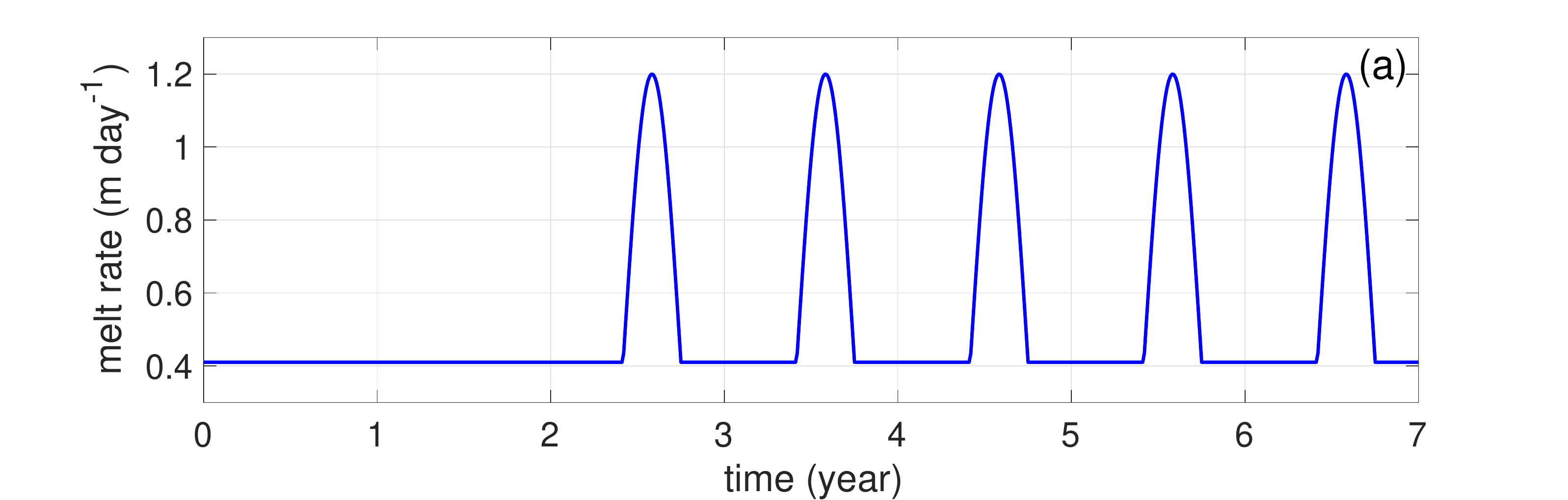}\\
    \includegraphics[width=0.65\textwidth]{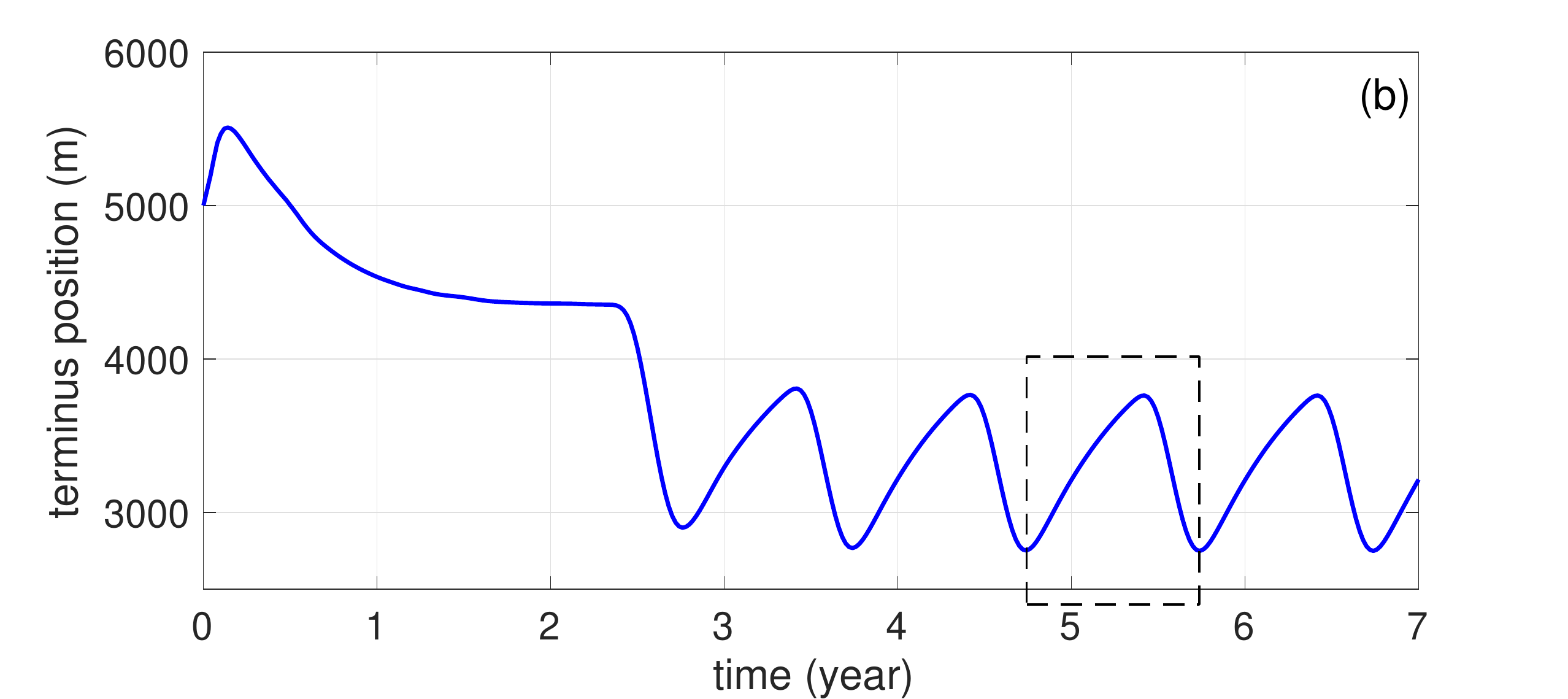}
    \caption{(a) Prescribed melt rate used as input for Experiment~1.
      (b) Computed variations in terminus position, with the one-year
      period $t \in [4.74, 5.74]$ highlighted by a dashed box.}
    % The shape of melt rate over a period of time. Initial melt rate is
    % considered as constant to reach the steady state surface
    % elevation. After reaching steady state, ice surface experience a
    % seasonal variation with a seasonal change of melt rate. In
    % seasonal change years, melting follows a sinusoidal pattern during
    % the summer months. Otherwise, a constant melt rate of
    % $0.41\,\mathrm{m}\,\mathrm{day}^{-1}$ is considered.}
    \label{fig:seasonal_meltrate_1to6yr}
  \end{center}
\end{figure}

This experiment has a total duration of 7~years and starts off during
the initial 2~years with a constant melt rate of $0.41\,\mathrm{m/day}$,
which is applied in order to generate a steady-state ice sheet
profile. After that, a periodic annual melt cycle with the following
time variation is imposed during each of the remaining 5~years
\begin{linenomath}
\begin{gather}
  \omega(t) = \left\{
    \begin{array}{ll}
      0.41, & \mathrm{if} \,\,0 \leqslant t \leqslant \frac{5}{12},\\[0.2cm]
      0.41 + (1.2 - 0.41) \sin(3 \pi t - 5\pi/4) , & \mathrm{if} \,\,
      \frac{5}{12} < t \leqslant \frac{9}{12},\\[0.2cm] 
      0.41, & \mathrm{if} \,\,\frac{9}{12} < t \leqslant 1,
    \end{array} \right.
  \label{eq:seasonal_melt}
\end{gather}
\end{linenomath}
(measured in m/day). The complete seven-year melting rate time series is
pictured in Fig.~\ref{fig:seasonal_meltrate_1to6yr}a.

The SSA velocity must be determined on a spatial domain extending from
$x=0$ to the grounded calving front, which is discretized on a regular
spatial grid with spacing of $\Delta x = 50\,\mathrm{m}$. Because the
terminus position varies continuously with time, $x_\mathrm{g}$ does not
necessarily coincide with a grid point; consequently, we perform the
velocity solve on the computational domain $[0, \, N_x\Delta x]$ where
$N_x = \left\lfloor {x_\mathrm{g}}/{\Delta x} \right\rfloor$ so that the
terminus point always lies in the interval
$N_x\Delta x \leqslant x_\mathrm{g} < (N_x + 1)\Delta x$.  A constant
velocity of $u_\mathrm{in} = 8000\,\mathrm{m/yr}$ is imposed on the
up-glacier boundary, whereas at the calving front we impose the discrete
form of
$\frac{\partial u}{\partial x} = \left(\frac{1}{4} A^{1/\alpha} (1 -
  \rho/\rho_\mathrm{w}) \rho g H \right)^\alpha$ at $x=x_\mathrm{g}$
using cubic extrapolation when the terminus lies outside the
computational domain.  The LSM solver employs a spatial domain of size
$[0, 10000\,\mathrm{m}]\times [0, 400\,\mathrm{m}]$ which is discretized
at a uniform $200 \times 100$ grid of points, ensuring that $\Delta x$
is the same for both LSM and SSA solvers and so the horizontal grid
locations interior to the ice region coincide. The coupled system is
integrated using a constant time step of
$\Delta t = 5 \times 10^{-4}\,\mathrm{yr}$.

%\begin{figure}[bthp]
%  \begin{center}
%    \includegraphics[width=0.7\textwidth]{figures/meltRateYearly.eps}
%    \caption{\centering  The annual cycle of the prescribed melt rate.
%      %      Melting parameterization at glacier bottom over a period of 1 year.
%    }
%    \label{fig:meltRateYearly}
%  \end{center}
%\end{figure}

%We consider the initial melt rate as constant
%$0.41\,\mathrm{m}\,\mathrm{day}^{-1}$ for the first two years to reach
%a steady state ice sheet profile before applying the seasonal
%variability \eqref{eq:seasonal_melt}. After two years, a seasonal
%change of melt rate  described in Eq.~\eqref{eq:seasonal_melt} is added
%using a sinusoidal pattern during five summer months. Otherwise, a
%constant melt rate of $0.41\,\mathrm{m}\,\mathrm{day}^{-1}$ is
%considered for other months (see Figure
%\ref{fig:seasonal_meltrate_1to6yr}). 

\subsubsection{Results of seasonal melt forcing}

% We apply the SSA--LSM ice model (see Section~\ref{sec:da_algorithm} or
% \cite{hossain2020modelling}) to track the ice surface interface and
% terminus for advancing and retreating ice sheets due to seasonal
% melting variability.

\begin{figure}[bthp]
  \begin{center}
    \includegraphics[width=0.8\textwidth]{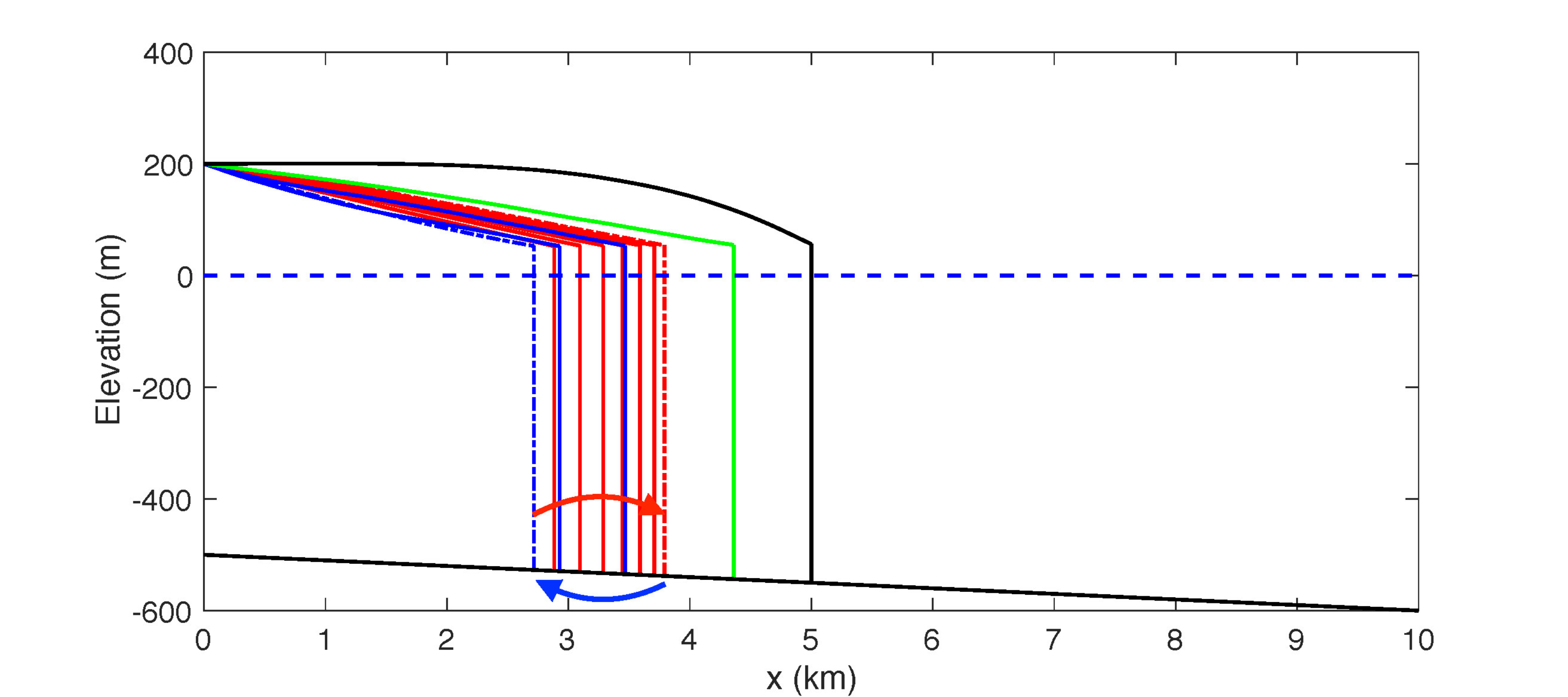}
    \caption{Variations in glacier surface profile over time for
      Experiment~1.  The black line depicts the initial ice surface from
      Eq.~\eqref{eq:surfaceseasonal} whereas the green line shows the
      surface after 2 years of constant melting. Following that, the ice
      experiences a seasonal cycle of advance and retreat in response to
      a periodically varying melt rate. The curves corresponding to
      advance (red) and retreat (blue) are also shown for a single
      annual cycle at every 0.1 years over the time interval
      $[4.74, 5.74]$.}
    \label{fig:icesurface_seasonal}
  \end{center}
\end{figure}

During the start-up period with a constant melt rate the ice sheet
experiences a period of relatively rapid advance, reaching its greatest
extent of $x_\mathrm{g} = 5509\,\mathrm{m}$ after 0.14~year. This is
simply an artificially induced transient behaviour arising from our
initial choice for the upstream glacier height.  Following that, the ice
sheet retreats and thins, gradually reaching a steady state after
roughly 2~years have elapsed (see
Fig.~\ref{fig:seasonal_meltrate_1to6yr}b).
% The simulation starts off with a constant melt rate
% $0.41\,\mathrm{m/day}$ during which time the ice sheet retreats and
% thins as expected, reaching a steady state after roughly 2 years.
Fig.~\ref{fig:icesurface_seasonal} depicts the corresponding initial ice
surface as a black curve, alongside the ``steady-state'' profile reached
after 2~years in green.

% \begin{figure}[bthp]
%   \begin{center}
%     \includegraphics[height=4.5cm, width=11cm]{figures/GLp-ssa-da2-Seasonal2_1}
%     \caption{\centering Variation of terminus position as a function of time.}
%     \label{fig:GLp}
%   \end{center}
% \end{figure}%

After the initial start-up period, the seasonal variations in melt rate
pictured in Fig.~\ref{fig:seasonal_meltrate_1to6yr}a begin to take
place, which causes the ice sheet to experience an annual cycle of
advance and retreat. This behaviour can be seen in
Fig.~\ref{fig:seasonal_meltrate_1to6yr}, where it is easy to connect
variations in terminus position with the annual variations in melt
rate. The terminus position clearly exhibits a periodic pattern with the
same period as the melt rate cycle, varying in extent between
$2749.7\,\mathrm{m}$ and $3807.8\,\mathrm{m}$ as measured from the
up-glacier domain boundary.

Throughout these annual melting cycles, the ice sheet reaches its
minimum extent at times $t=2.76$, 3.74, 4.74, 5.74 and 6.74 years, with
the successive maxima occurring at $t=3.42$, 4.42, 5.42 and 6.42
years. The red and blue lines in Fig.~\ref{fig:icesurface_seasonal}
picture the advancing/retreating ice profiles at 10 equally-spaced times
separated by 0.1 year during the annual period corresponding to
$t \in [4.74, 5.74]$ (highlighted in the dashed box in
Fig.~\ref{fig:seasonal_meltrate_1to6yr}b).  The red curves show the
glacier while advancing and the blue curves during retreat, with the two
outermost dash-dot lines representing the maximum and minimum glacier
extent.  The speed of advance is significantly slower than that for
retreat, which is reflected in the fact that more red lines are required
to capture the advancing stage. This cycle of slow advance followed by
fast retreat is also clearly captured in the plot of terminus position
in Fig.~\ref{fig:seasonal_meltrate_1to6yr}b where the leading edge of
each peak is considerably shallower than the trailing edge. More
specifically, the advance from minimum to maximum extent requires
approximately 0.68 year, while the subsequent retreat lasts only 0.32
year so that the glacier retreats roughly twice as fast as it advances.

%%%%%%%%%%%%%%%%%%%%%%%%%%%%%%%%%%%%%%%%%%%%%%%%%%%%%%%%%%%%%%%%%%
\subsection{Experiment 2: Seasonal ice dynamics with data
  assimilation}
\label{sec:exp_design}

The second experiment is a modified version of Experiment~1 that applies
a ``pseudo-random'' variation in melt rate to more realistically mimic
actual weather. We also incorporate the data assimilation portion of the
algorithm to illustrate how exploiting observational data within the
solution updates can improve accuracy.
%
% We perform a twin experiment to evaluate the performance of the data
% assimilation algorithm.
%
% In this section we first describe the synthetic reference experiment
% that will be used to assess the performance of the data assimilation
% framework. 
%
 
% \subsubsection{Reference simulation}
\subsubsection{Experimental design -- Truth run and synthetic
  observations}
\label{sec:exp2-truth}

\begin{figure}[bthp]
  \begin{center}
    \includegraphics[width=0.65\textwidth]{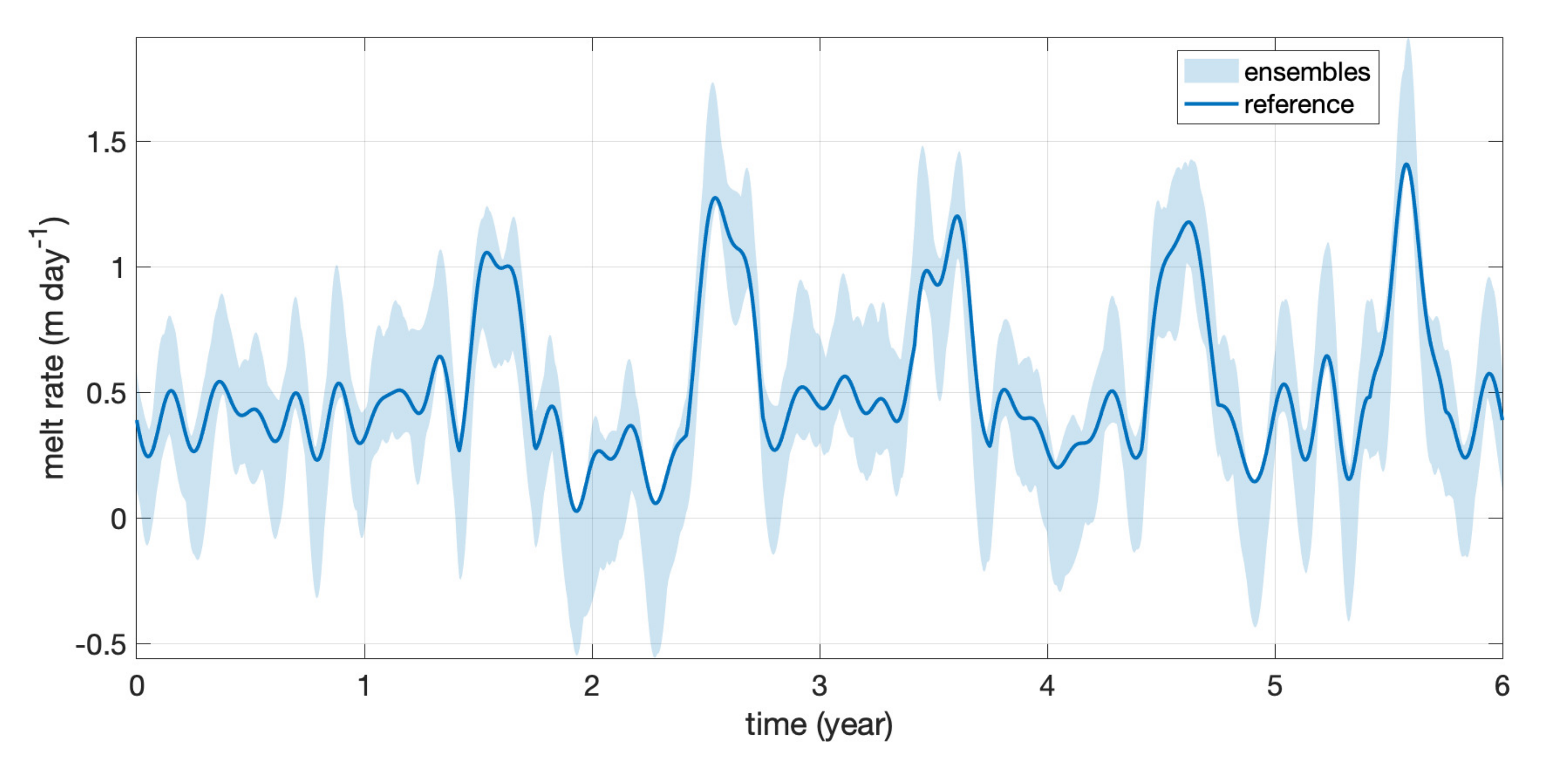}
    \caption{Melt rate time series for the data assimilation
      Experiment~2.  The true/reference melt rate is shown as a solid
      blue line and the shaded light blue region depicts the spread in
      melt rates represented by the $N_e=50$ ensemble members.}
    \label{fig:enkfmeltrate0}
  \end{center}
\end{figure}

We start by generating a model solution using the SSA--LSM algorithm
with initial and boundary conditions from Experiment~1, where
Eq.~\eqref{eq:surfaceseasonal} is taken as the ``true'' or reference
state for the initial glacier surface, again with $h_0=150\,\mathrm{m}$
so that the up-glacier boundary is fixed at $200\,\mathrm{m}$ above the
sea level.
% However, we modify the idealized seasonal melt rate from
% Fig.~\ref{fig:seasonal_meltrate_1to6yr}, using it as a baseline to
% which is added a random perturbation built from a linear superposition
% of 41 sinusoids with a given sequence of wavenumbers and
% randomly-chosen amplitudes and phase shifts.
However, we modify the melt rate forcing from
Fig.~\ref{fig:seasonal_meltrate_1to6yr}a by applying the constant melt
rate $0.41\,\mathrm{m}\,\mathrm{day}^{-1}$ in year~1 only, followed by a
seasonal rate that tracks Eq.~\eqref{eq:seasonal_melt} as a baseline but
adds random variability.
% using a linear superposition of 41 sinusoidal waves with differing
% amplitudes, wavenumbers and phase change.
Because of ``pseudo-random'' nature of this melt rate, the glacier never
reaches a steady state, which is why we only apply the constant melt
rate for a single year (compared with 2 years for Experiment~1). The
constant melt rate during year~1 also allows us to observe the terminus
dynamics in the absence of seasonal melt rate variations.
% I think it may be best here to write each sinusoid as
% $A \sin(2\pi f t + \phi)$ and then state the values of f.
Random variability is incorporated through a linear superposition of 41
Fourier modes of the form $a_i \sin(2\pi f_i\, t + \phi_i)$, with
frequencies $f_i=\frac{i}{6}$ for $i=0, \dots, 40$ (where the
denominator 6 represents the number of years in this experiment).  The
amplitude and phase shift for each mode are both chosen randomly from
normal distributions, with $a$ having mean 0 and standard deviation
0.025, and $\phi$ with mean 0 and standard deviation $2\pi$. This
procedure is used to generate a unique melt rate distribution for the
truth run as well as for each ensemble member.  In Fig.~\ref{fig:enkfmeltrate0},
the true (or reference) melt rate is displayed as a thick blue curve,
while the melt rates corresponding to the randomly-generated ensembles
lie within the shaded region surrounding this curve. The simulation
based on the reference melt rate is considered the ``truth run'' and is
used to measure the quality of our data assimilation estimates.

We next describe the process for generating synthetic observations in
the data assimilation process.  Based on the truth run described above,
we generate randomly perturbed observations by sampling ice surface and
terminus position from Gaussian distributions with means equal to the
truth values and standard deviations
$\sigma_{h}^{\mathrm{obs}} = 10\,\mathrm{m}$ and
$\sigma_{x_\mathrm{g}}^{\mathrm{obs}} = 50\,\mathrm{m}$
respectively. Note that this introduces vertical errors in the surface
elevation whereas the terminus error is in the horizontal
direction. Synthetic observations are generated from the truth run at 10
equally-spaced time intervals each year (for a total of $N_y=50$
observations) with samples of surface elevation taken at equidistant
points along the flow line with separation distance of
$800\,\mathrm{m}$.
% The model is run for 6 years with a time step $\Delta t = 5 \times
% 10^{-4}\,\mathrm{yr}$ and LSM solver runs on uniformly spaced
% grid size $200 \times 100$. 

\subsubsection{Initial ensembles}
\label{sec:Initial_Ensembles}

\begin{figure}[bthp]
  \begin{center}
    \includegraphics[width=0.6\textwidth]{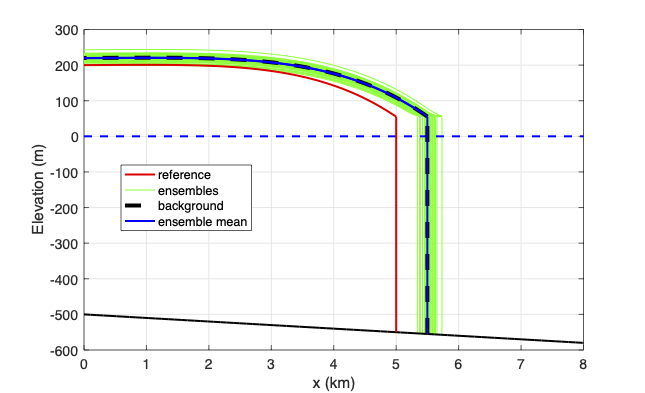}
    \caption{For Experiment~2, the reference or true ice
      surface profile is shown in red, the background profile as a
      dashed black line, the ensemble profiles by green lines, and the
      ensemble mean in blue.}
    % Black dash represents the background, greens are initial
    % ensembles, and Blue represents the ensemble average.
    \label{fig:enkfinitial}
  \end{center}
\end{figure}

There is no well-accepted rule for generating the $N_e$ initial
ensembles and so we have chosen to employ a common
approach~\cite{bonan2017data} that adds Gaussian noise to the background
state for the ice surface height.  We also introduce an additional
horizontal offset by scaling the reference terminus position with a
constant factor, replacing $x_\mathrm{g}$ with $\lambda x_\mathrm{g}$ in
Eq.~\eqref{eq:surfaceseasonal}.
% \begin{linenomath}
%   \begin{gather}
%     h(x) = h_0\left(1 - \left(\frac{x}{\beta
%           x_\mathrm{g}}\right)^4\right) - (1 - \rho/\rho_w) b(x),
%     \quad 0 \leqslant x \leqslant \beta x_\mathrm{g}  
%     \label{eq:surfaceseasonal_background}
%   \end{gather}
% \end{linenomath}
In our EnKF simulations, the ensembles have a mean initial terminus
position larger than that for the truth corresponding to $\lambda=1.1$.
%both smaller and larger than the truth (corresponding
%to $\lambda=0.9$ and $1.1$) but here we only present results for the case
%$\lambda=1.1$.
To generate ensemble values for terminus position $x_\mathrm{g}^{(i)}$, we add
Gaussian noise to the scaled terminus position $\lambda x_\mathrm{g}$
that is sampled from the normal distribution
$\mathcal{N}(0,\,\sigma_{\lambda x_\mathrm{g}}^2)$ with standard deviation
$\sigma_{\lambda x_\mathrm{g}} = 50\,\mathrm{m}$. 
% variance is \sigma^2
Similarly, the up-glacier surface elevations $h_0^{(i)}$ are
supplemented with noise sampled from $\mathcal{N}(0,\,\sigma_{h_0}^{2})$
with $\sigma_{h_0}=10\,\mathrm{m}$. The initial ice surface profile for each ensemble is
then generated by replacing \eqref{eq:surfaceseasonal} with
% to the background surface Eq.~\eqref{eq:surfaceseasonal_background} we
% can define the surface for each ensembles:
\begin{linenomath}
  \begin{gather*} 
    h^{(i)}(x) = h_{0}^{(i)}\left( 1 - \left( x/{x_\mathrm{g}^{(i)}}
      \right)^4 \right) - (1 - \rho/\rho_\mathrm{w}) b(x) \quad \text{for} \; 0
    \leqslant x \leqslant x_\mathrm{g}^{(i)},
    \label{eq:surfaceseasonal_ensembles}
  \end{gather*}
\end{linenomath}
for ensembles $i=1,2,\dots,N_e$. For this experiment, we generate
$N_e=50$ ensemble members (consistent with the EnKF approach
in~\cite{gillet2020assimilation}) for which our initial ice surface
profiles and reference/background states are pictured in
Fig.~\ref{fig:enkfinitial}.  Note that each ensemble is forced using a
unique pseudo-random melt rate time series that is different from that
of the reference melt rate (see
Fig.~\ref{fig:enkfmeltrate0}). Furthermore, unlike for the truth run,
the surface elevations $h_0^{(i)}$ at the up-glacier boundary are not
fixed, but instead can vary over time.

\subsubsection{Results of ETKF simulation with seasonal melt forcing}

Using the initial background and ensemble states, we apply the
forecast--analysis--reinitialization sequence outlined in the
SSA--LSM--ETKF algorithm in Section~\ref{sec:da_algorithm}. For
comparison purposes, this experiment is repeated using the same initial
background states but without data assimilation. Both simulations are
subjected to the same pseudo-random melt rate time series described in
Section~\ref{sec:exp2-truth} and pictured in
Fig.~\ref{fig:enkfmeltrate0}. Note that the ice surface at the
up-glacier boundary for this comparison run can also change with time,
similar to the data assimilation case.

\begin{figure}[bthp]
  \begin{center}
    \includegraphics[width=0.45\textwidth]{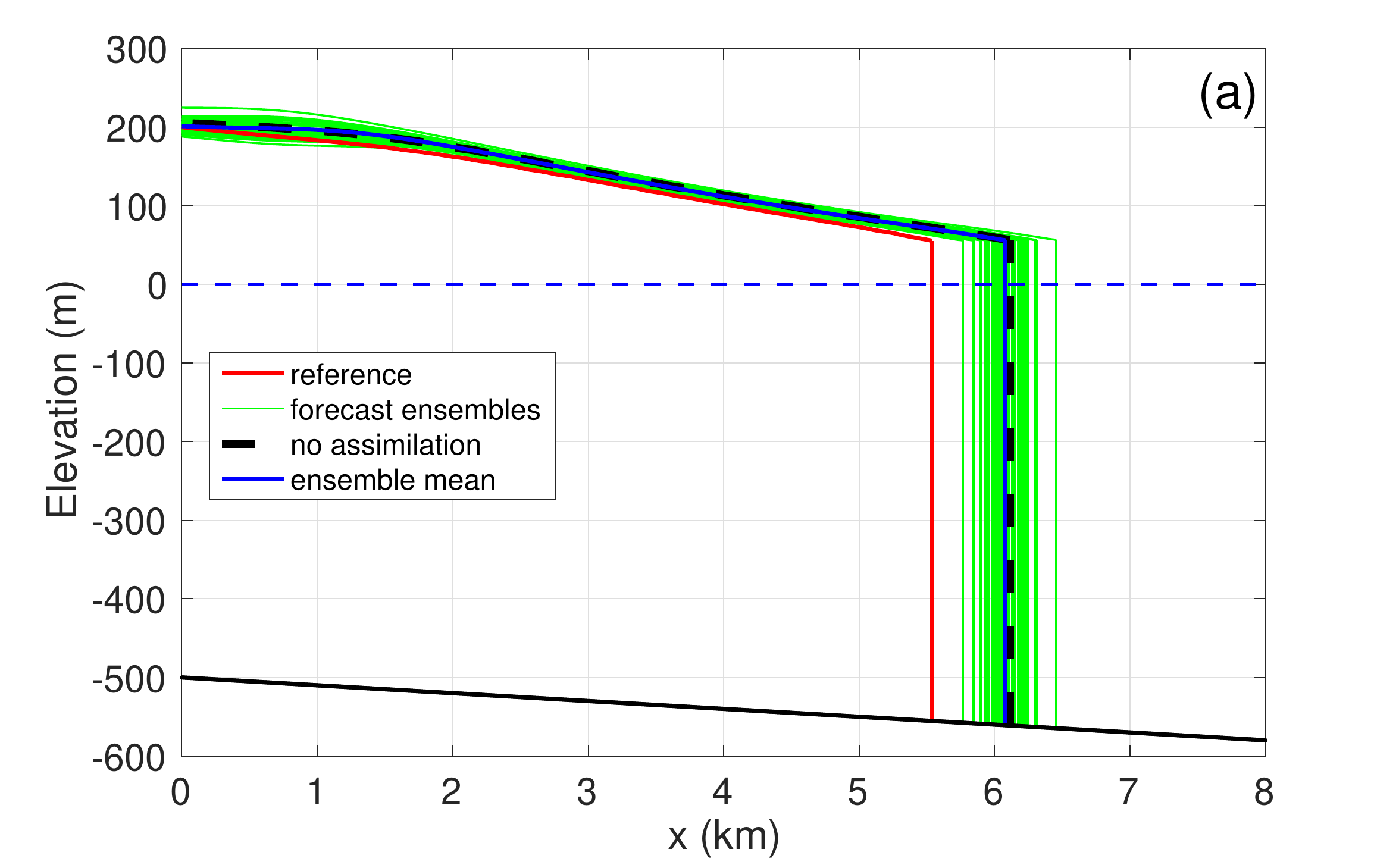}\\
    \includegraphics[width=0.45\textwidth]{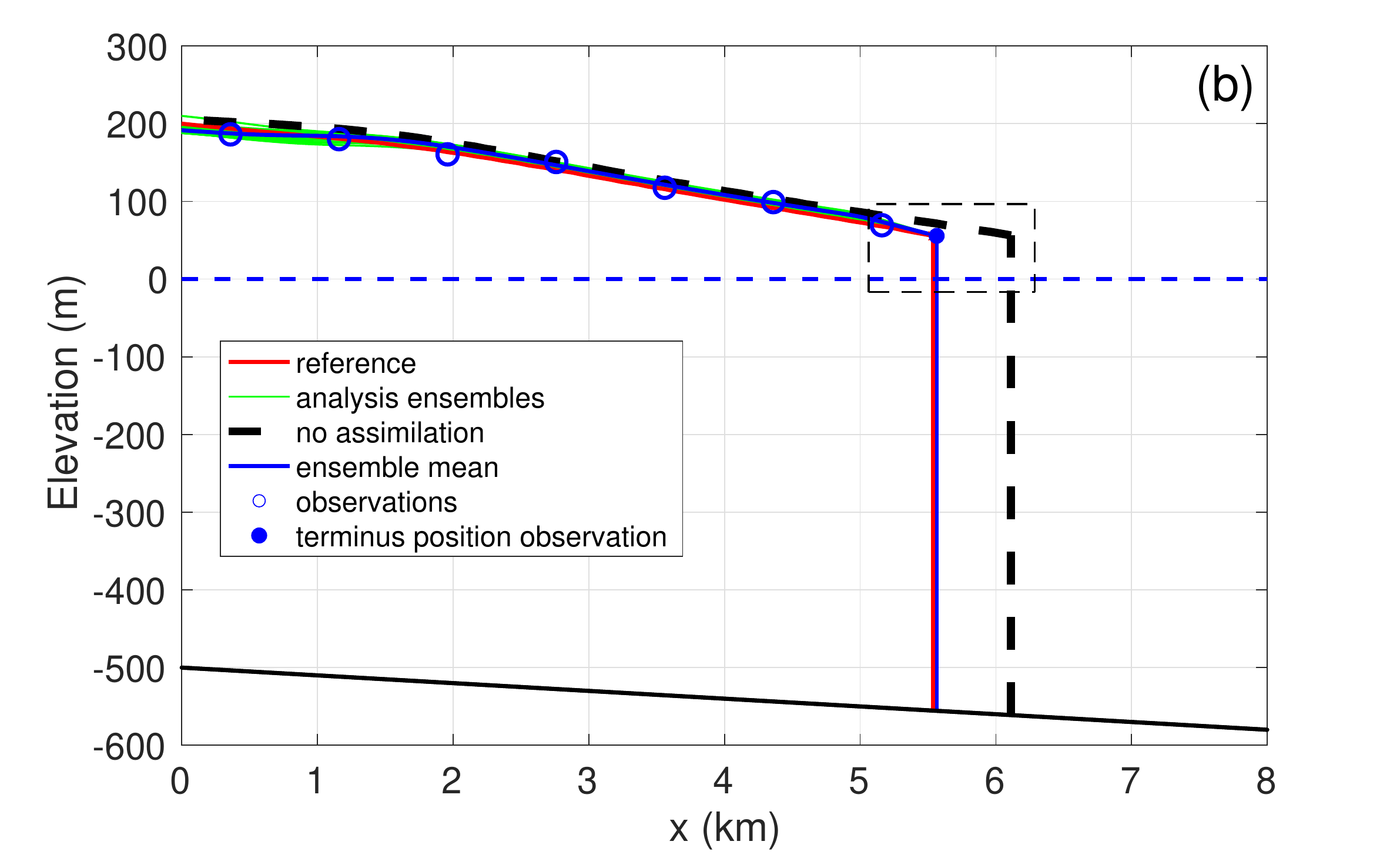}
    \includegraphics[width=0.45\textwidth]{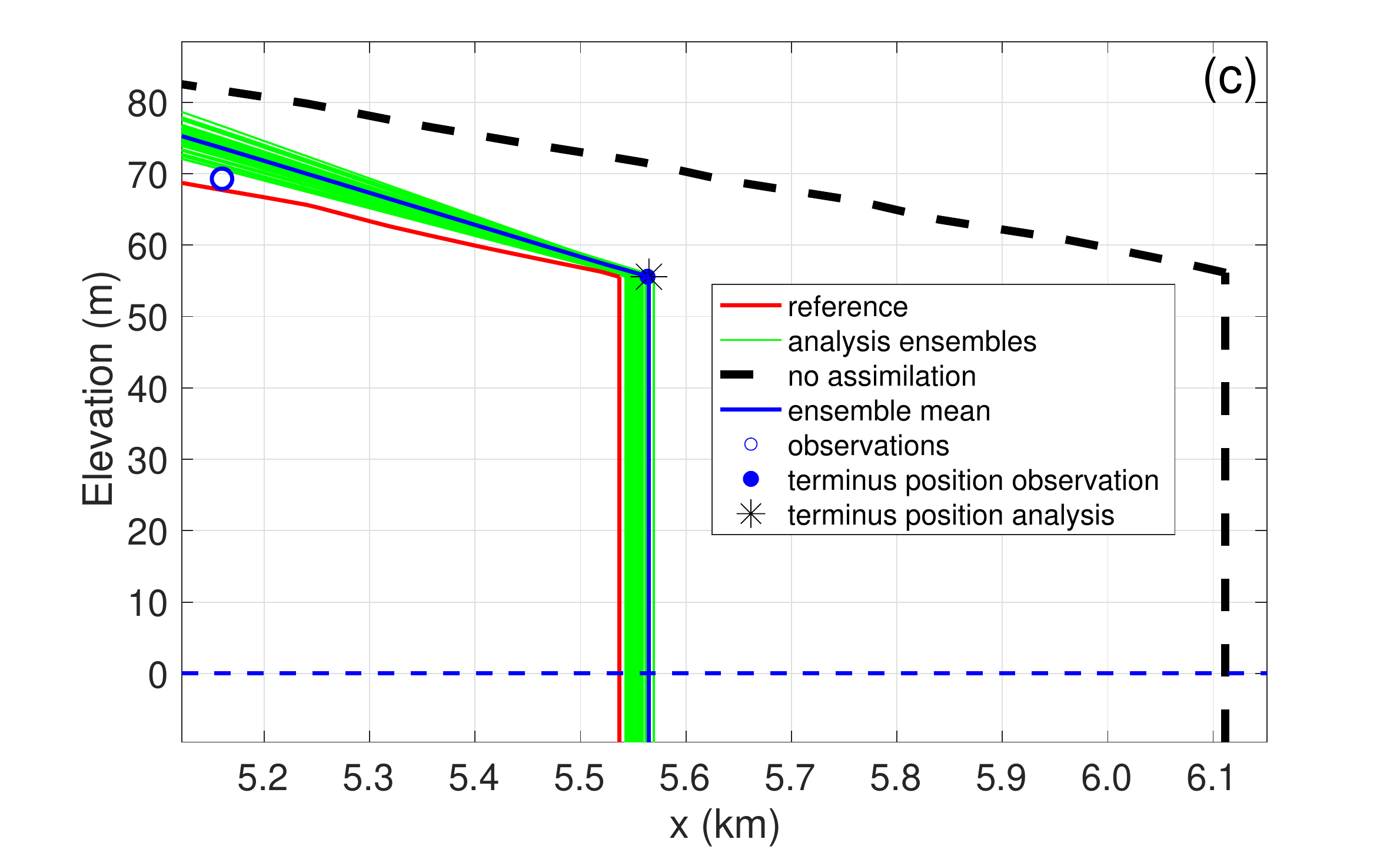}
    \caption{Solution for Experiment~2 at time $t=0.1$~year.  (a) ETKF
      forecast (before data assimilation) compared with the
      reference. (b) ETKF analysis (after data assimilation) compared
      with reference. (c) A zoomed-in view of the boxed region near the
      terminus highlighted in plot (b). The reference or true ice stream
      is shown in red, the background as a black dashed line, ensembles
      in green, and the ensemble average in blue.}
    \label{fig:seasonal0_1}
  \end{center}
\end{figure}

Between times $t=0$ and $0.1$~yr, we apply the forecast step in
Section~\ref{sec:da_algorithm} to predict the ice surface and terminus
position for the $N_e$ ensembles, with the reference state being used to
generate the first set of synthetic observations. The assimilation
results from $t=0.1$ are summarised in Fig.~\ref{fig:seasonal0_1}, where
the ensembles of forecast terminus position exhibit a spread between
$5765\,\mathrm{m}$ and $6455\,\mathrm{m}$ (refer to
Fig.~\ref{fig:seasonal0_1}a).  The analysis step updates the solution as
pictured in Fig.~\ref{fig:seasonal0_1}b, where the terminus observation
of $5563.5\,\mathrm{m}$ causes the analysis ensembles to shift into a
much tighter range lying between $5542\,\mathrm{m}$ and
$5570\,\mathrm{m}$.  The mean value of the forecast ensembles is
$6080\,\mathrm{m}$, and after the data assimilation step the analysis
ensemble mean is $5565\,\mathrm{m}$.
%This result should be compared to that obtained by evolving the background state without data assimilation, for which the terminus position is $6111\,\mathrm{m}$. 
Considering that the reference terminus solution is
$5537\,\mathrm{m}$, the data assimilation process is clearly successful
in moving the ensembles significantly closer to the desired target
solution.
% The analysis ensemble mean of terminus position moves toward the
% reference terminus position and reaches to $5565\,\mathrm{m}$ from the
% forecast ensemble mean terminus position $6080\,\mathrm{m}$ where the
% reference terminus position is $5537\,\mathrm{m}$ and background state
% terminus position without assimilation it is $6111\,\mathrm{m}$.
Likewise, we observe at the up-glacier boundary that the ETKF process
also narrows the range of ensemble analysis heights from the forecast
range $[188,225]\,\mathrm{m}$ to that of the analysis $[188, 210]\,\mathrm{m}$, with the
reference/truth value being fixed at $200\,\mathrm{m}$.

\begin{figure}[bthp]
  \begin{center}
    \includegraphics[width=0.65\textwidth]{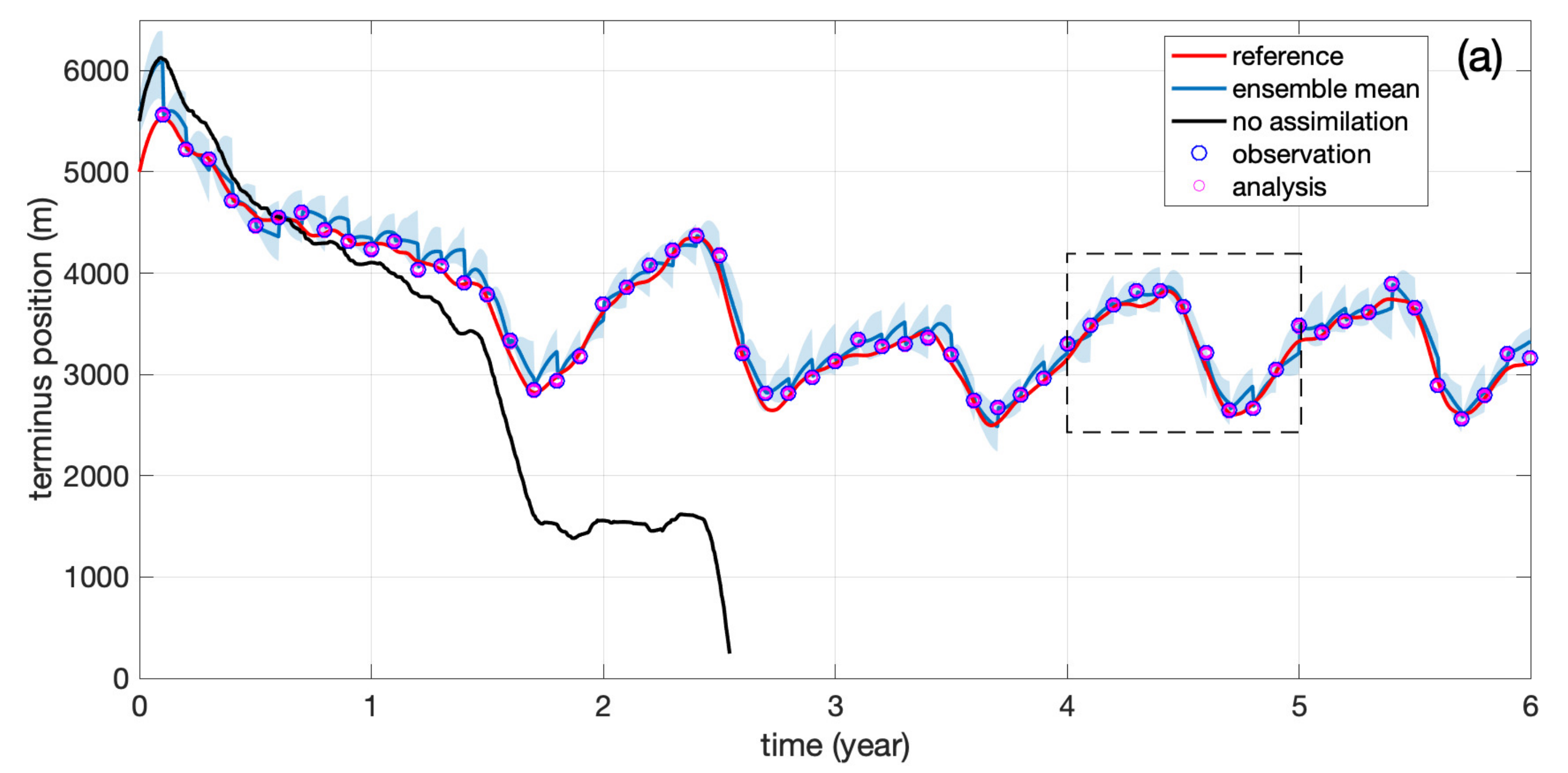}
     \includegraphics[width=0.65\textwidth]{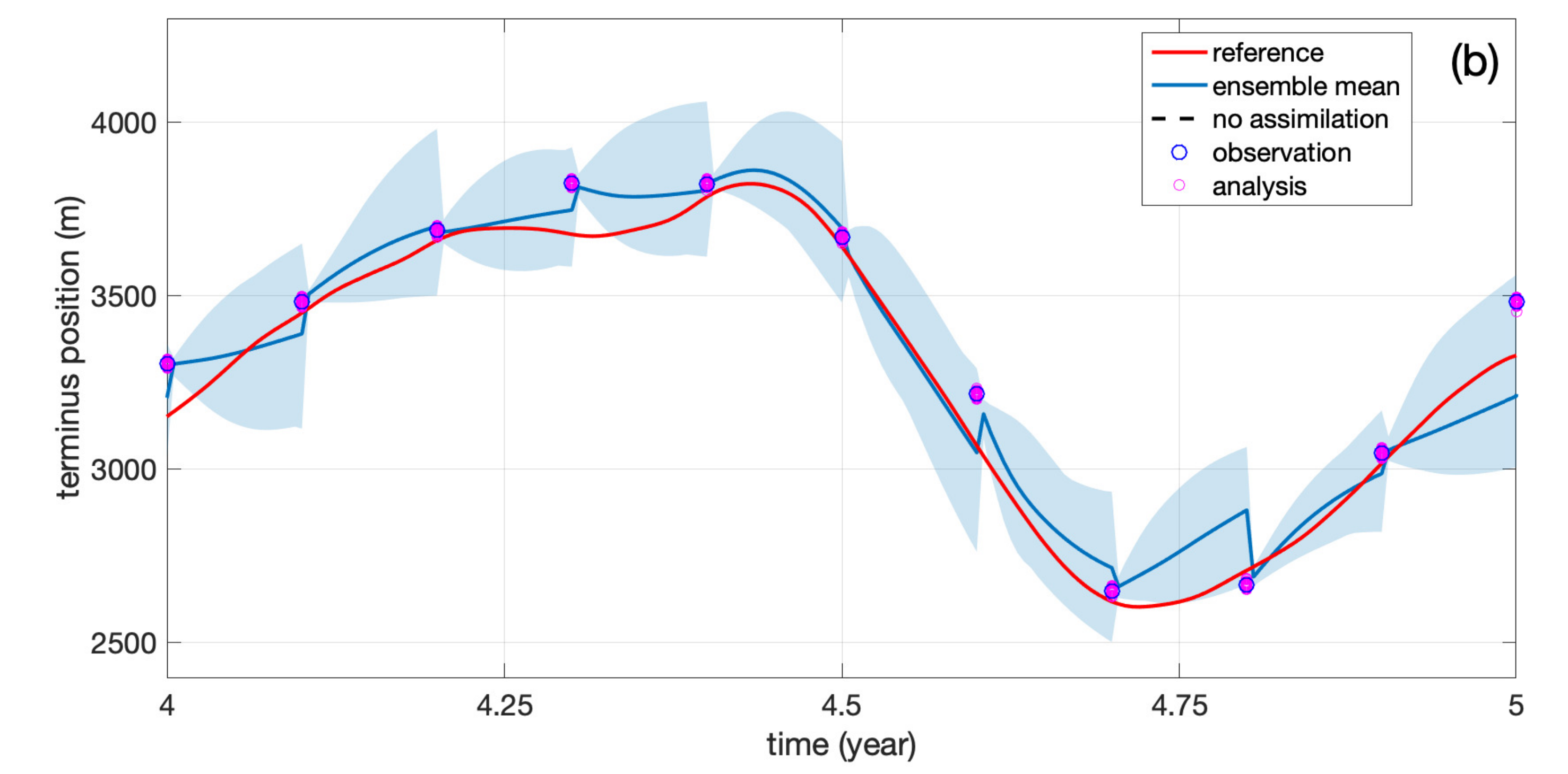}
     \caption{(a) Time series of terminus positions for
       Experiment~2.  The reference solution is shown as a red line, the
       ensemble mean in blue, and the ensemble spread by the shaded
       area.  Observation points are available every 0.1 years and shown
       as circular points.  The simulation without data assimilation is
       shown as a black line.  (b) Zoomed-in view of the highlighted
       area of the terminus plot from plot (a).  By focusing in on the
       time period $[4,5]\;$years, it is much easier to recognize the
       observation and analysis points and the effect of assimilating
       observations into the analysis ensembles.}
    % This figure shows the state, observations, and $N_e = 50$ ETKF
    % ensemble members for grounding lines. Observations of grounding
    % line are assimilated for 10-times a year and a forecast is made
    % for further 0.1 years.} 
    \label{fig:etkf_groundinglines}
  \end{center}
\end{figure}

We continue applying the forecast--analysis--reinitialization sequence
over the full 6-year time period with observational data assimilated
every 0.1~yr. After the first full year of the simulation, seasonal melt
rate variations begin that drive cycles of ice sheet advance and retreat
similar to those observed in Experiment~1. Based on the plot of terminus
position in Fig.~\ref{fig:etkf_groundinglines}a we observe similar
behaviour to the results from Fig.~\ref{fig:seasonal_meltrate_1to6yr},
where there is no seasonal melt rate variability during the initial
start-up period, after which the terminus position again experiences an
approximately periodic pattern of slow advance and rapid retreat.  A
particularly striking comparison is offered with the simulation without
data assimilation which suggests the ice sheet retreats completely after
roughly 2.55 years, in stark contrast with the clear cycle of
advance/retreat suggested by the data. This is likely due to the fact
that the up-glacier boundary height is fixed throughout the entire
simulation, whereas the data assimilation algorithm allows the
up-glacier height to adjust adaptively with the ensembles in each
analysis step.
% A particularly striking result is that in the simulation without the
% ETKF step the ice sheet retreats completely after roughly 2.55 years,  
% Whereas the ice geometry recovers and terminus position exhibits an
% annual periodic cycle when observations are assimilated. 

% Grounding line is moving from a minimum of 2800 m from our up-glacier
% boundary to reach a maximum of 3800 m during the period of season
% variation melt rate.

%\begin{figure}[bthp]
%  \begin{center}
 %   \includegraphics[width=0.55\textwidth]{figures/etkf_magnified4_5_sept}
 %   \caption{Zoomed-in view of the highlighted area of the
 %     terminus plot from Experiment~2 in
 %     Fig.~\ref{fig:etkf_groundinglines}.  By focusing in on the time
 %     period $[4,5]\;$years, it is much easier to recognize the
 %     observation and analysis points and the effect of assimilating
 %     observations into the analysis ensembles.}
%    \label{fig:etkf_groundinglines_magnified}
%  \end{center}
%\end{figure}
\begin{figure}[bthp]
  \begin{center}
    \includegraphics[width=0.55\textwidth]{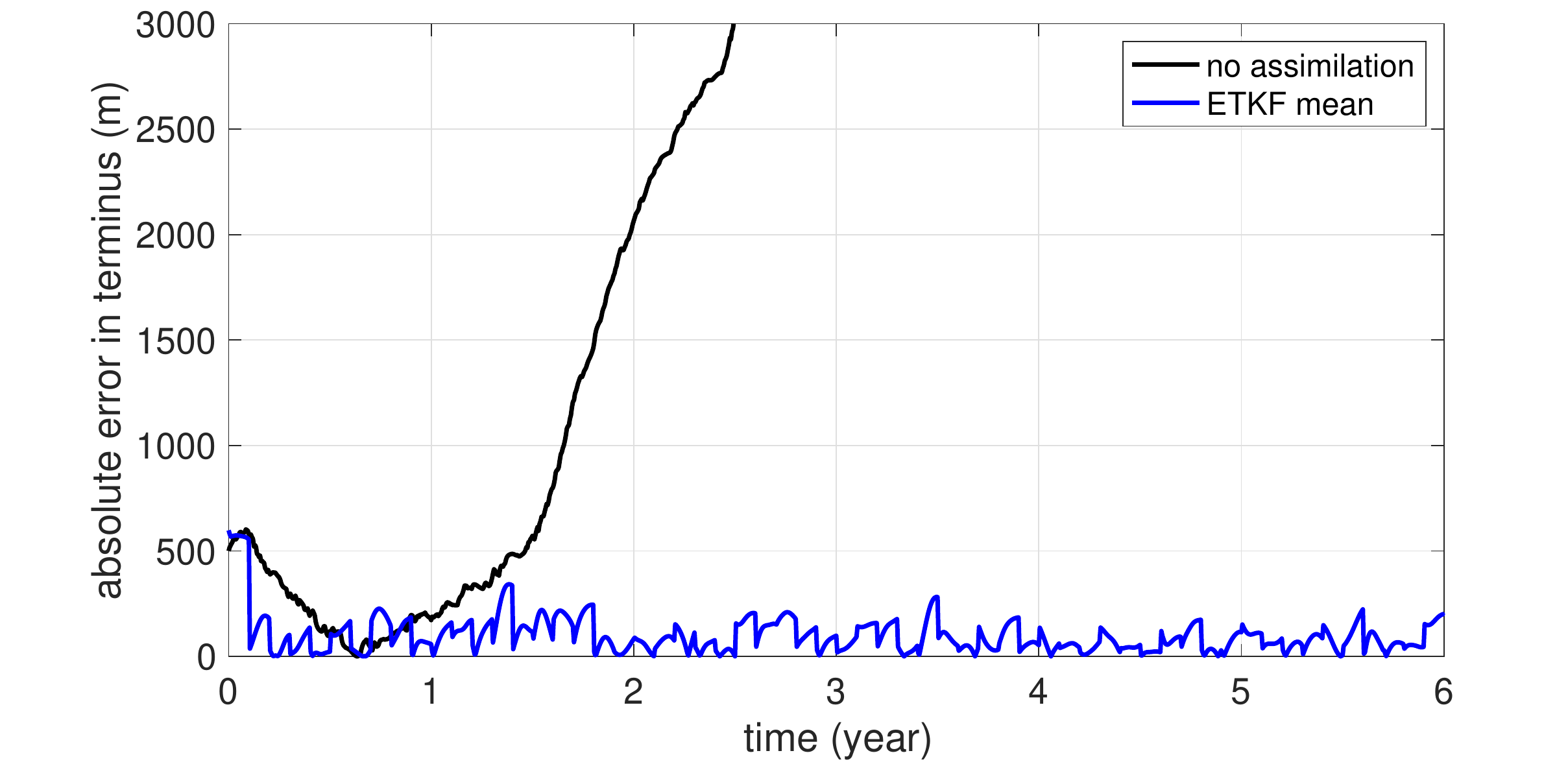}
    \caption{Mean absolute error in terminus position for
      Experiment~2, with and without data assimilation.}
    \label{fig:etkf_groundinglines_abs_error}
  \end{center}
\end{figure} 

A zoomed-in view of the terminus plot corresponding to the time period
$[4,5]\,\mathrm{yr}$ (highlighted in
Fig.~\ref{fig:etkf_groundinglines}a) is depicted in 
Fig.~\ref{fig:etkf_groundinglines}b. This plot shows more clearly how
the terminus estimate is improved in each data assimilation step and
drawn back towards the reference solution.  The ETKF procedure provides
consistent estimates in the sense that the reference terminus position
always lies within the shaded area representing the spread of
ensembles. Fig.~\ref{fig:etkf_groundinglines_abs_error} depicts the
evolution of the mean error in terminus position and demonstrates that
the ETKF generates reliable forecasts up to the next assimilation time.
% For example, within the very first analysis step at time $t=0.1$, the
% mean error decreases from $546$~m to $28\,\mathrm{m}$. This
% well-controlled nature of the errors in the ETKF algorithm should be
% compared to those without data assimilation, where the errors
% accumulate rapidly in time as the glacier eventually retreats beyond
% the up-glacier boundary.
For example, in the first analysis step at time $t=0.1$ the absolute
error in terminus position of the ensemble mean is reduced to
$28\,\mathrm{m}$ from the uncorrected value of $546\,\mathrm{m}$, with a
similar reduction in every data assimilation step.  This well-controlled
nature of the ETKF errors should be compared to the errors without data
assimilation, shown as black curves in
Figs.~\ref{fig:etkf_groundinglines}a
and~\ref{fig:etkf_groundinglines_abs_error}.  Without data assimilation,
there is an initial transient reduction in error over the initial time
interval $[0,0.635]$, after which the error grows steadily and
eventually becomes so large that the glacier retreats completely beyond
the up-glacier boundary.
      
%%%%%%%%%%%%%%%%%%%%%%%%%%%%%%%%%%%%%%%%%%%%%%%%%%%%%%%%%%%%%%%%%%
\subsection{Experiment 3: Helheim Glacier in Greenland}

The final experiment applies our coupled data assimilation algorithm to
simulate the motion of Helheim Glacier which is one of the largest
outlet glaciers of the Greenland Ice Sheet.  The simulated glacier is
evolved in time with the SSA--LSM model and remotely-sensed observations
of terminus position and surface elevation are incorporated at various
times over the period May~2001 to August~2006 using the ETKF.
% In this case, we have access to actual data for bottom topography and
% remotely-sensed observations of surface elevation and surface
% velocities.
Helheim is a tidewater-terminating glacier that experienced a rapid
retreat during the early 2000s followed by a subsequent re-advance
(see~\cite{howat_etal_2008}, for example).  This period of dramatic ice
loss, ice-flow acceleration, and glacier thinning is thought to have
resulted from an increase in air temperatures that triggered surface
melt-induced thinning and increased basal lubrication, along with
increased ocean temperatures that destabilized the glacier
terminus~\cite{joughin2008ice, Straneo_etal_2011}.  This behaviour has
led to the hypothesis that the dynamic mass loss from the Greenland Ice
Sheet seen over the past two decades has been triggered largely by
perturbations occurring at the terminus of outlet glaciers such as
Helheim~\cite{straneo2013challenges}, and it seems likely that Helheim
Glacier is set for another period of dramatic
retreat~\cite{Williams_etal_2021}. With this in mind, dynamic models
that accurately track the terminus position of rapidly changing glaciers
are of particular importance for the entire ice
sheet~\cite{catania2020future}.  

% We assume that the true initial condition including the glacier
% surface and bed elevation along the near-terminus flow line of our
% simulation is known from the observed profile.

% We next describe the configuration of the reference simulation that
% will be used to assess the performance of the algorithm.

\subsubsection{Experimental design}
\begin{figure}[bthp]
  \begin{center}
    \includegraphics[width=0.47\textwidth]{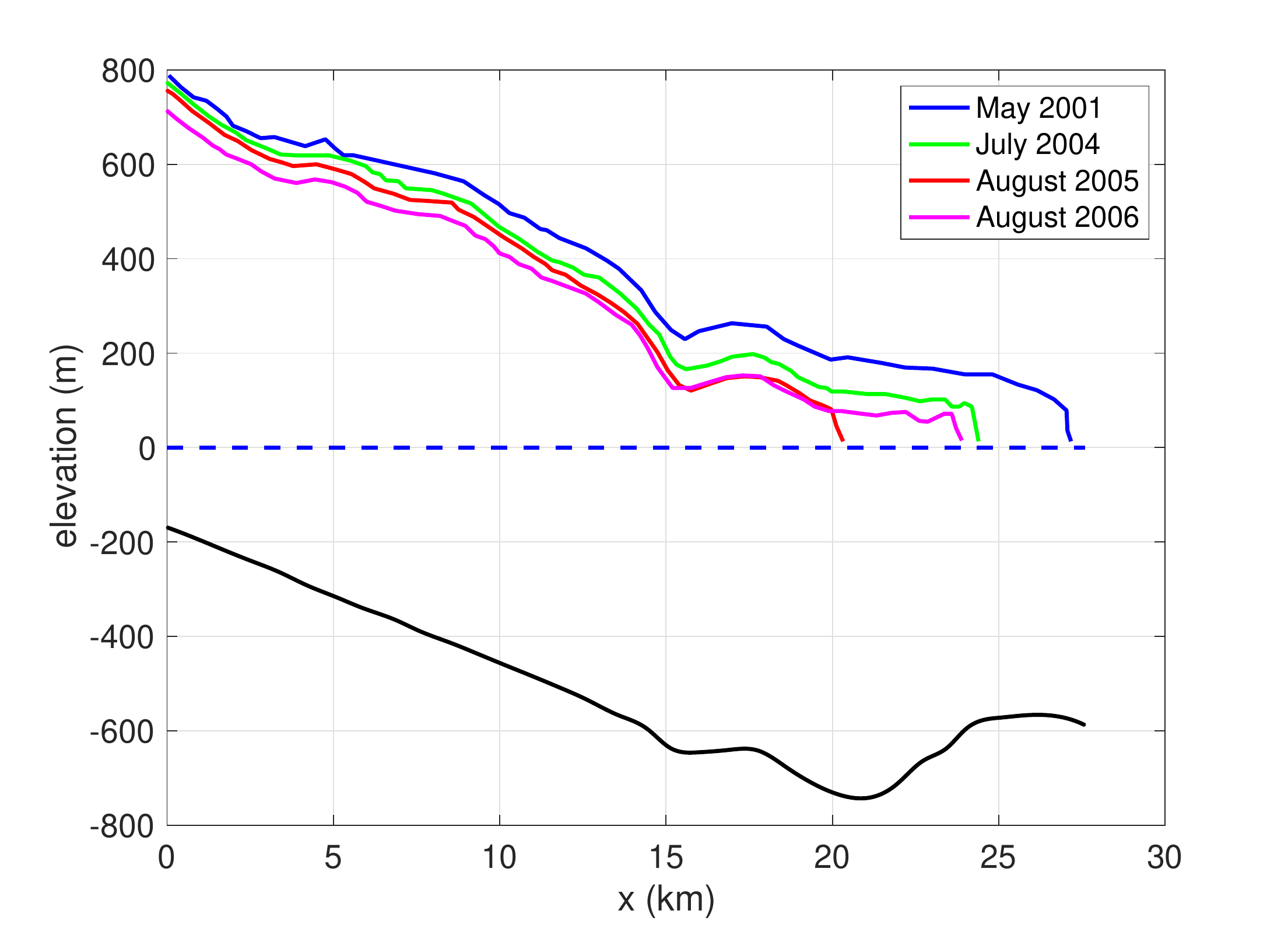}
    \caption{Bedrock and ice surface profiles for
      Helheim Glacier, which have been extracted digitally from Nick et
      al.~\cite[Fig.~1]{nick2009large} using WebPlotDigitizer
      (\url{https://automeris.io/WebPlotDigitizer}).}   
    \label{fig:helheim_real_data}
  \end{center}
\end{figure}

We now describe the configuration of the numerical experiment that will
be used to assess the performance of our SSA--LSM--ETKF algorithm.
% The bedrock profile for Helheim Glacier along the flow line is taken
% from~\cite{nick2009large} and depicted as a black curve at the bottom
% of Fig.~\ref{fig:helheim_real_data}.  The initial height of the upper
% ice--air surface is the glacier elevation recorded in
% May~2001~\cite{nick2009large} and plotted as a blue line in
% Fig.~\ref{fig:helheim_real_data}.
The initial ice geometry is taken from Nick et al.~\cite{nick2009large}
and is based on measured data of glacier surface and bedrock elevations
along the near-terminus flow line (see
Fig.~\ref{fig:helheim_real_data}).  The horizontal location of the
up-glacier boundary is chosen to be $28\,\mathrm{km}$ from the initial
terminus. Velocity observations from~\cite{nick2009large} suggest that
horizontal surface velocity at the up-glacier boundary $x=0$ varies
between 4.0 and $5.2\,\mathrm{km/yr}$ during the period May~2001 to
August~2006. For simplicity, we choose an up-glacier velocity of
$u_\mathrm{in}$ = $4.5\,\mathrm{km/yr}$ that lies near the middle of
this range and is held constant throughout the experiment, which means
that the total depth-integrated flux through the up-glacier boundary is
constant between observations.

The computational domain for the SSA velocity solve extends from the
up-glacier boundary at $x=0$ to the grounded calving front at
$x=x_\mathrm{g}$.  The left-hand boundary condition is
$u(0)=u_\mathrm{in}=4.5\,\mathrm{km/yr}$ and a slight modification of
the flotation condition \eqref{eq:ssa3} is imposed on the right
\begin{linenomath}
  \begin{gather*}
    \frac{\partial u(x_\mathrm{g})}{\partial x} = C_F A \left(
      {\textstyle\frac{1}{4}} \rho (1-\rho/\rho_\mathrm{w}) g
      H(x_\mathrm{g}) \right)^n,
    \label{eq:per_cf}
  \end{gather*}
\end{linenomath}
in order to capture the rapid discharge of ice into the
ocean. Following~\cite{nick2009large} we introduce an extra scaling
factor $C_F$ into Eq.~\eqref{eq:picard_bc2}, which is a constant front
stress perturbation coefficient, with $C_F=1$ denoting an unperturbed
calving front and $C_F>1$ indicating an increased longitudinal strain
rate at the terminus position due to rapid discharge (our specific
choice for $C_F$ is explained in Section~\ref{sec:Initial_Ensembles2}).
% which is used to apply the front-stress
% perturbations~\cite{brondex2017sensitivity}. 

To capture ice loss due to surface melting we impose a linear variation
of net melting rate with distance, increasing from 0\,m/yr on the
upstream boundary to 40\,m/yr at the initial terminus $x=28\,\mathrm{km}$.
%\begin{linenomath}
%\begin{gather*}
 % \omega(x) = \begin{cases}
%    4, & \text{if $0 \leqslant x \leqslant 10$}, \\[0.2cm]
%    \displaystyle 4+\frac{36(x-10)}{x_g-10},
%    & \text{if $10 < x \leqslant x_g$}.
%  \end{cases}
%\end{gather*}
%\end{linenomath}
The basal friction is assigned the constant value $C=3.0 \times
10^{-2}\,\mathrm{MPa\,m^{-1/3}\,yr^{1/3}}$, while all other physical
parameters are chosen the same as in other experiments (see
Table~\ref{tab:params}).
% The flow parameter $A$ is the rate factor relating to the rheological
% softness of the ice and is set to $5.6\times10^{-17}\,
% \mathrm{Pa^{-3}\,yr^{-1}}$, corresponding to an ice temperature of
% $-5^\circ$C.
The horizontal and vertical grid spacings for the SSA and LSM
discretization are $\Delta x=0.2\,\mathrm{km}$ and $\Delta z=10
\,\mathrm{m}$ respectively, and the time step is $\Delta t =
5 \times 10^{-3}\,\mathrm{yr}$.
% and LSM solver runs on uniformly spaced grid size $200 \times 100$.
Finally, within the data assimilation process we incorporate
observations of terminus position and surface height (the latter sampled
at $2\,\mathrm{km}$ intervals), both of which are extracted from
the~\cite{nick2009large} profiles shown in
Fig.~\ref{fig:helheim_real_data}. These data are then perturbed by
random noise sampled from a Gaussian distribution, as explained in the
next section.

\subsubsection{Initial ensembles}
\label{sec:Initial_Ensembles2}

\begin{figure}[bthp]
  \begin{center}
    \includegraphics[width=0.55\textwidth]{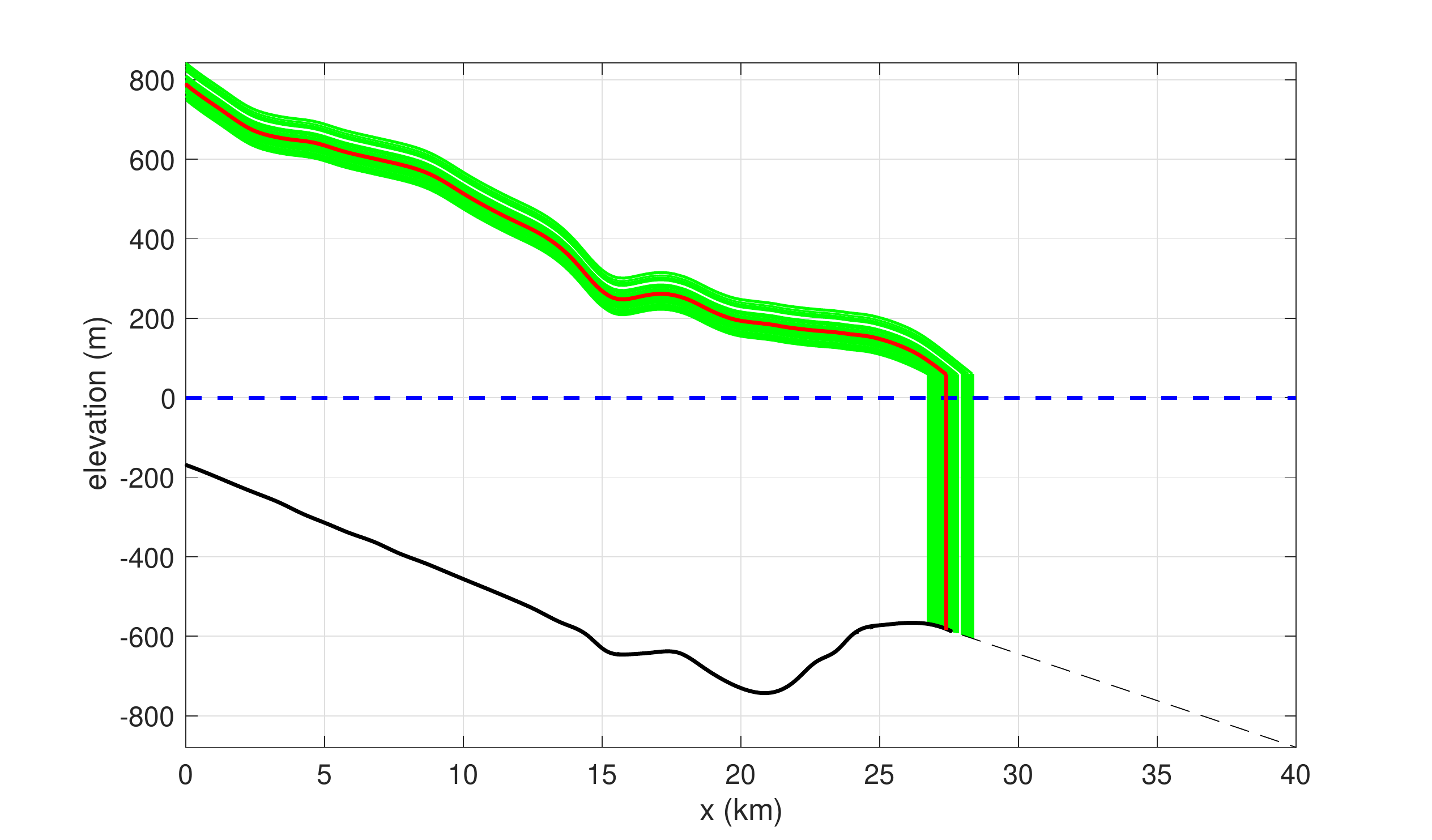}
    \caption{Initial data for Experiment~3, with the reference or true
      ice stream in red and the initial ensembles in green.}
    \label{fig:enkfinitial_nick}
  \end{center}
\end{figure}

To determine the initial ensembles for Experiment~3, we take the initial
background state to be the ice profile observed on May~2001 (see
Fig.~\ref{fig:helheim_real_data}). Ensembles are generated by adding
Gaussian noise from $\mathcal{N}(0,\,\sigma^{2})$ to the surface height
and choosing terminus positions that satisfy the flotation criteria,
which is repeated for $N_e=50$ ensemble members with
$\sigma=20\,\mathrm{m}$ (see Fig.~\ref{fig:enkfinitial_nick}). The
flotation criterion is imposed at the terminus for all ensemble members
as follows: for any member with surface height lying above the
background state, the surface is extrapolated linearly up to the
position where the flotation criterion is satisfied; if the surface lies
below the background state, then the ice sheet is truncated at the point
where flotation criteria is satisfied. Each ensemble is simulated using
a different value of the stress perturbation coefficient chosen from the
range $[1.5,\,3.5)$, with $C_F = 1.5 + \frac{2}{N_e}(i-1)$ for ensembles
numbered $i=1, 2, \dots, N_e$. Note that this range includes the value
$C_F=2.8$ used by Nick et al.~\cite{nick2009large} in their modelling
studies of Helheim Glacier.

\subsubsection{Tracking ice thickness and terminus for Helheim Glacier}

Observational data for surface elevation and terminus position are
extracted for this experiment from the measured surface height profiles
for Helheim Glacier at three analysis times: 
July~2004, August~2005 and August~2006 (refer to
Fig.~\ref{fig:helheim_real_data}).
% for 10 equally spaced time in a year from the reference run.
For surface elevation we choose equidistant observation points
separated by a distance of $2\,\mathrm{km}$ along the flow direction,
and add uncorrelated Gaussian noise with zero mean and standard
deviation $\sigma_h^\mathrm{obs}=20\,\mathrm{m}$. A similar process is
used to generate observations of the terminus position except that a
standard deviation of $\sigma_{x_\mathrm{g}}^\mathrm{obs} =
200\,\mathrm{m}$ is used instead.

\begin{figure}[bthp]
  \begin{center}
    \includegraphics[width=0.49\textwidth]{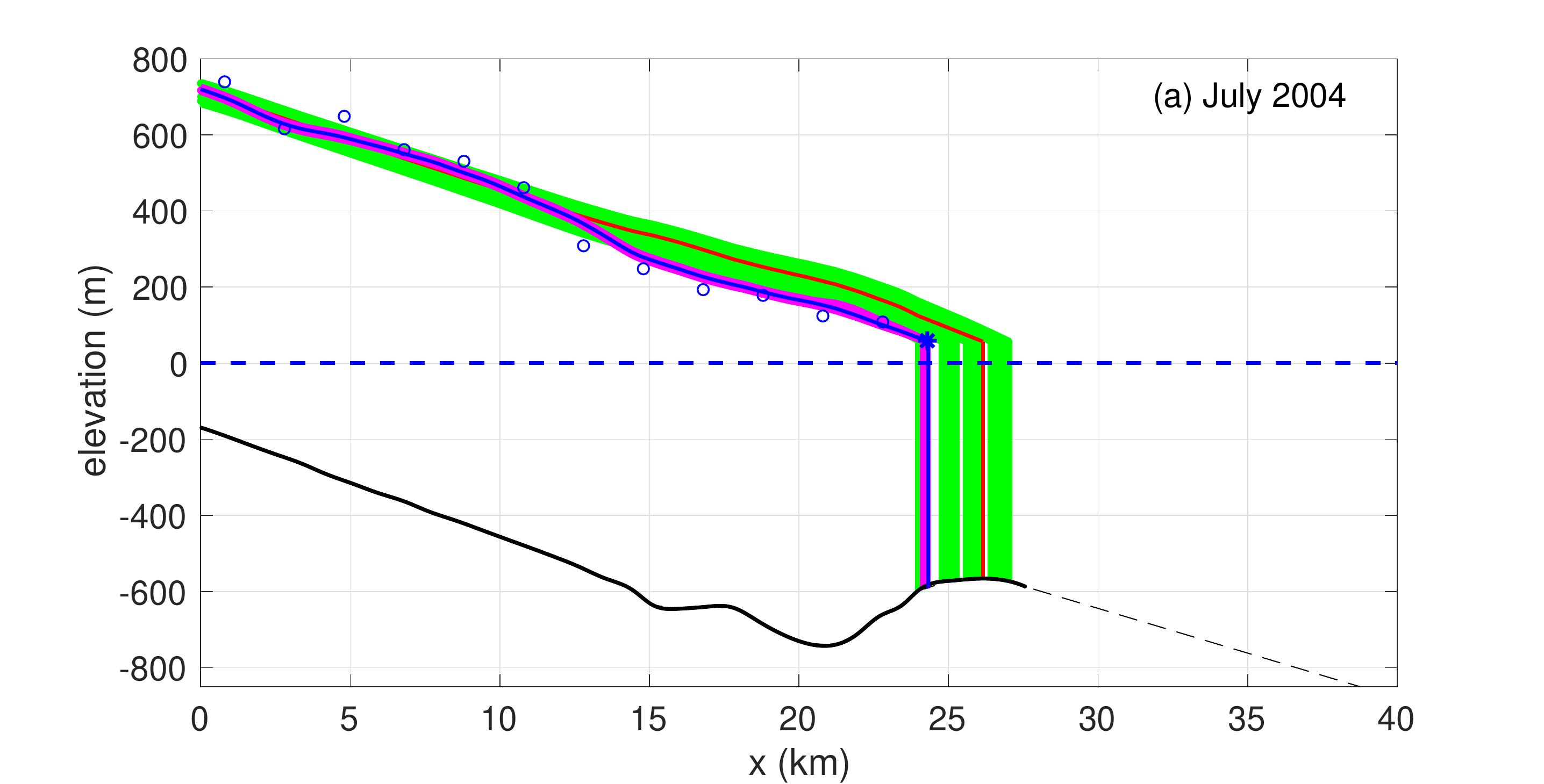}
    \includegraphics[width=0.49\textwidth]{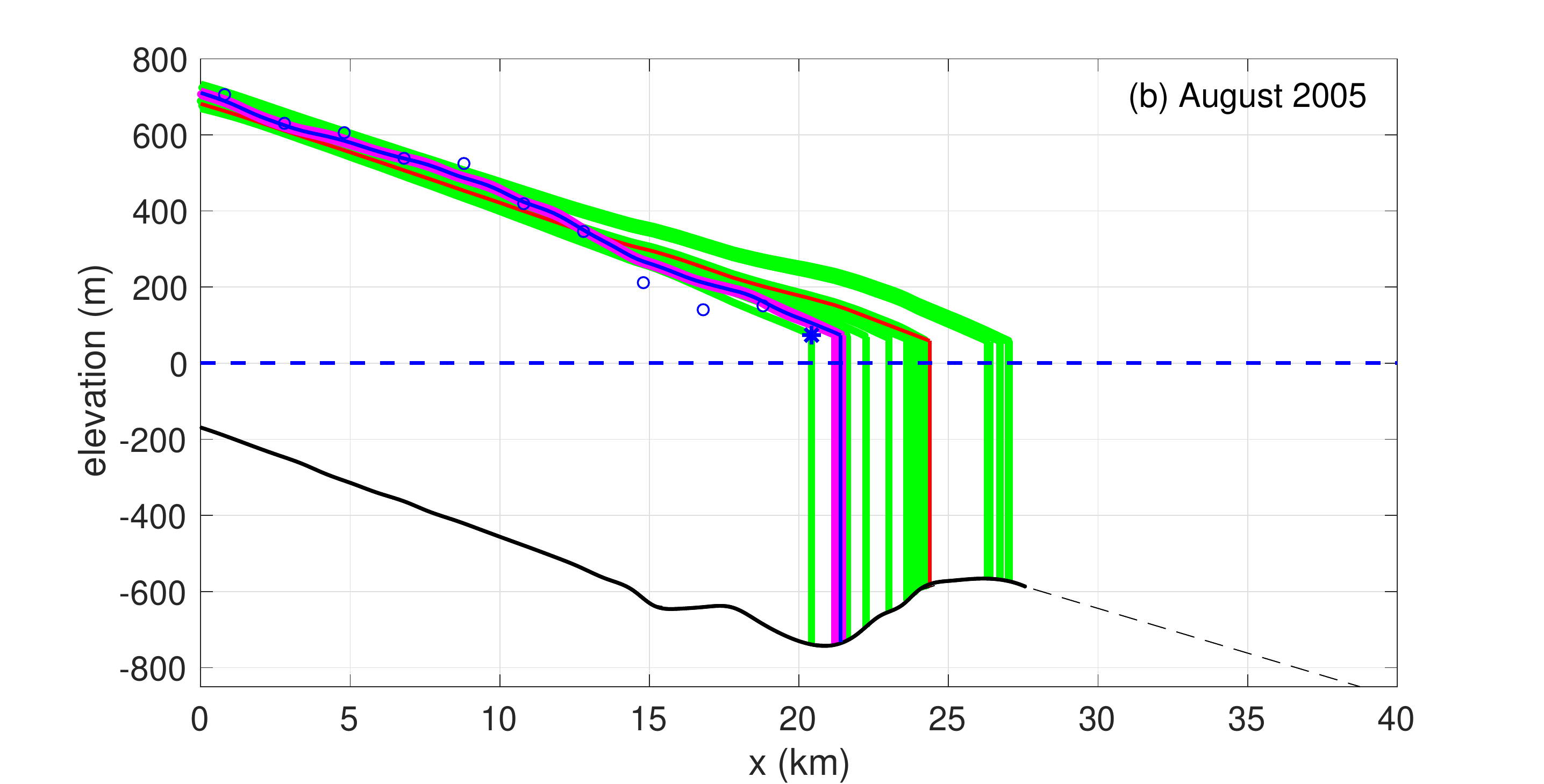}\\
    \includegraphics[width=0.49\textwidth]{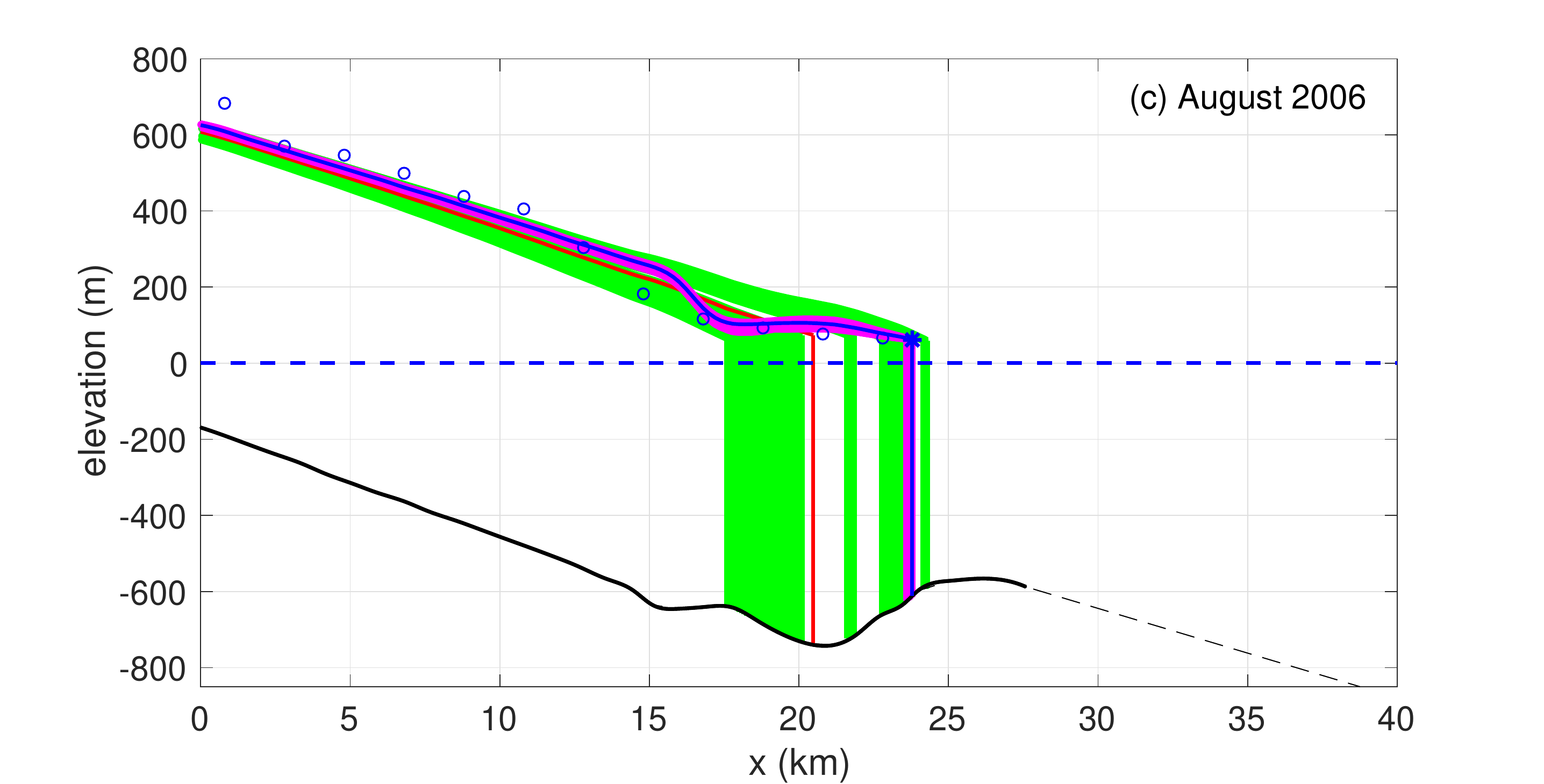}
    \caption{Forecast and analysis ensembles for Experiment~3 at the
      three analysis times when observations are assimilated (July~2004,
      August~2005, August~2006).  The green and magenta curves represent
      forecast and analysis ensembles respectively, and the
      corresponding means are plotted in red and blue. The `$\circ$'
      symbol represents ice surface observation points and `$\ast$' the
      terminus observations.}
    \label{fig:helheim_etkf}
  \end{center}
\end{figure}

Using this data, we may then compute the ice sheet motion by applying
the SSA--LSM--ETKF algorithm from Section~\ref{sec:da_algorithm} and
performing the forecast--analysis--reinitialization process at the three
analysis times. The plots in Figs.~\ref{fig:helheim_etkf}a--c display
the forecast and analysis ensembles on each of these three dates.  At
the initial time May~2001, we apply the forecast step to predict the ice
surface and terminus using $N_e$ ensembles. After integrating the SSA
model to July~2004, the terminus positions in the forecast ensembles all
lie within the range $[24.0,27.0]\,\mathrm{km}$.  The corresponding mean
terminus position is $26.2\,\mathrm{km}$
% The mean terminus position is close to the observed terminus at this
% time with a difference of $22\,\mathrm{m}$.
which lies $1.9\,\mathrm{km}$ downstream of the observation
point. Moving to the ice surface profiles, note that the ensemble
elevations encompass the observation point at the up-glacier boundary,
with a forecast ranging between $[682.4, 737.0]\,\mathrm{m}$ at $x=0$
and with a mean value of $713.7\,\mathrm{m}$.

The algorithm proceeds next to incorporate observations from July~2004
along with the forecast to determine analysis ensembles for ice surface
and terminus position. It is clear from Fig.~\ref{fig:helheim_etkf}a
that the ETKF process has contracted the analysis ensembles (in magenta)
into a tighter band much closer to the observations. The analysis
terminus position lies within the spread of the forecast ensembles,
although it has retreated further than most of the forecasts predict. We
note also that the forecast ensembles diverge between upward- and
downward-sloping portions of the bedrock.  At this stage, the
level set functions are rebuilt using the FMM and the SSA model
integration continues to the next analysis step.

\begin{figure}[bthp]
  \begin{center}
    \includegraphics[width=0.7\textwidth]{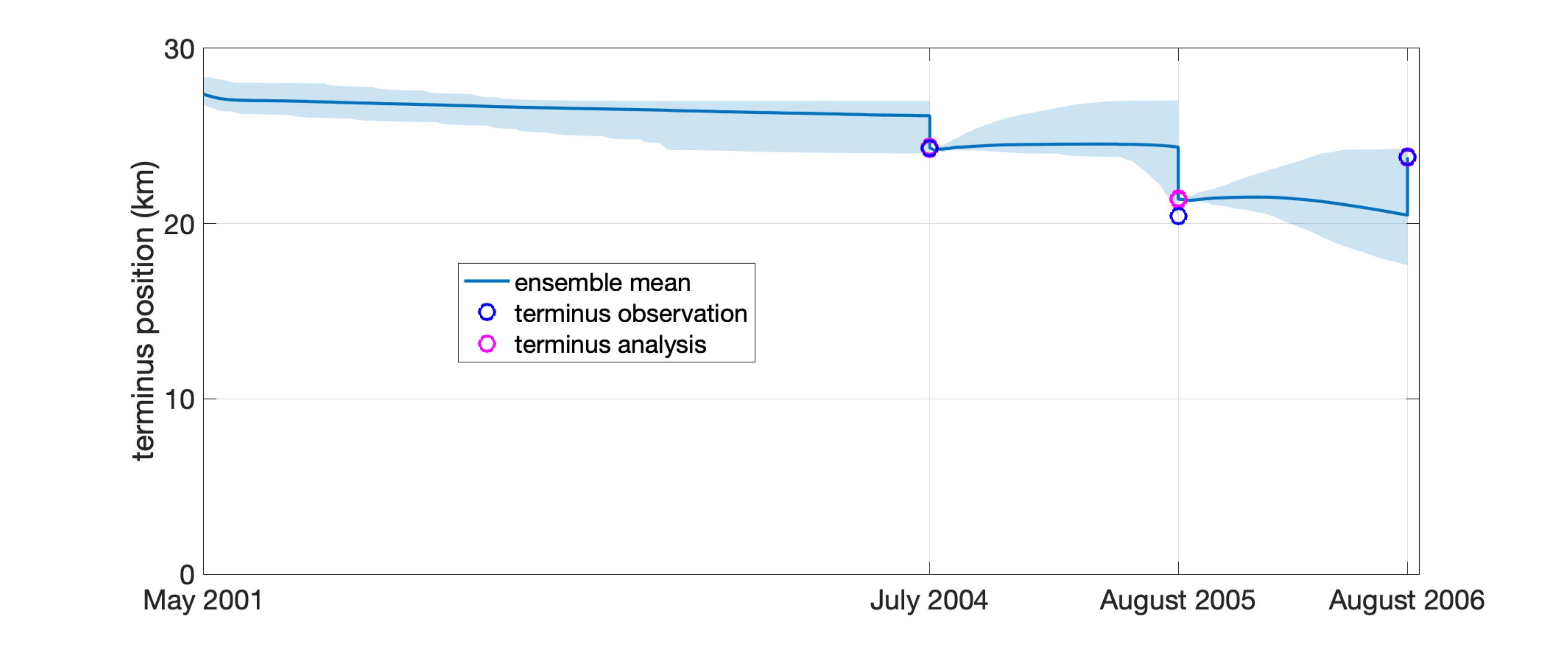}
    \caption{Terminus positions for Experiment~3, showing the range
      covered by the ensembles (light blue shaded region), ensemble mean
      (blue curve) and observation/analysis points (circles).}
    \label{fig:helheim_terminus}
  \end{center}
\end{figure}

Upon reaching August~2005 (refer to Fig.~\ref{fig:helheim_etkf}b) the
ensembles exhibit a noticeably larger spread than in the previous step,
with the terminus positions ranging between $[20.4, 27.0]\,\mathrm{km}$.
This demonstrates the inherent uncertainty in the forecast resulting
from the nonlinearity of the problem and the sensitivity to slight
changes in the bedrock slope at the terminating position as well as the
parameter choice $C_{F}$ (which represents buttressing at the terminus).
Twelve of the ensembles advance over the bedrock bump and are
concentrated on the downward-sloping portion of the bed between
$[26.3, 27.0]\,\mathrm{km}$, whereas the remaining 38 ensembles retreat
(some quite rapidly) and terminate along the upward-sloping portion of
the bed. The migration over time of the terminus positions can be seen
in Fig.~\ref{fig:helheim_terminus}, where the mean forecast terminus
position on August~2005 is $24.4\,\mathrm{km}$.  However, the observed
position indicates a dramatic retreat that places the terminus near
$x=20.4\,\mathrm{km}$, which coincides with the most rapidly retreating
forecasts (at one extreme of the ensemble distribution).  This unusually
rapid rate of glacier retreat during the 13 months spanning July 2004 to
August 2005 (when the observed terminus retreats by $4.1\,\mathrm{km}$,
compared to the three-year period from May 2001 to July 2004 when it
retreats only $2.8\,\mathrm{km}$) has been the subject of much
discussion in the literature (see~\cite{howat_etal_2008, joughin2008ice,
  Straneo_etal_2011, straneo2013challenges}).

The data assimilation step succeeds in shifting the simulated terminus
back toward the observed retreated location and thereby reducing
uncertainty in the solution (see
Fig.~\ref{fig:helheim_terminus}). Focusing again on the up-glacier
boundary in Fig.~\ref{fig:helheim_etkf}b, the simulations yield
elevations that are centred around the observation which is similar to
the July~2004 analysis step.  At lower glacier elevations, the ETKF
correction step lowers the ice surface as suggested by the observations.
We again reinitialize all ensembles and continue integrating to the
final observation point.

% At Aug~2006, all ensembles retreated faster from Aug~2005 these
% ensembles loses elevation and GRL become shorter. On the other hand,
% observation shows GRL advanced from Aug~2005. We again apply ETKF
% to reinitialize ensembles for the next forecast elevation and GRL
% position.  

Between August~2005 and August~2006, Fig.~\ref{fig:helheim_etkf}c shows
that the forecast again exhibits a strong sensitivity to the initial
condition and parameter choice, with some ensembles advancing up the
upward sloping bed and others retreating up the downward sloping
bed. Observations indicate that Helheim Glacier actually stabilized and
re-advanced during this time period, likely in response to an
anomalously cold year over Greenland~\cite{joughin2008ice}. The data
assimilation update, by combining measured observations with the
numerical forecast, is shown to address the forecast uncertainties and
improve forecasting skill by reinitializing the ice sheet geometry.
% the observations clearly show that the glacier advances (to the right)
% along the upward sloping bed, which is consistent with 2006 being an
% anomalously cold year for Greenland~\cite{joughin2008ice}. On the
% other hand, over the same time period forecast ensembles advances and
% retreats and analysis ensembles stay within the range of forecast
% ensembles.  All plots in Fig.~\ref{fig:helheim_etkf} show that
% terminus positions of ensembles are highly sensitive at this tidewater
% outlet glacier.

% some comments here regarding the point of da in such situations ...!
% how it copes with spatiotemporal chaotic systems?  Moreover, data
% assimilation could be an essential approach to reinitialize the model.
 
\begin{figure}[bthp]
  \begin{center}
    \includegraphics[width=0.7\textwidth]{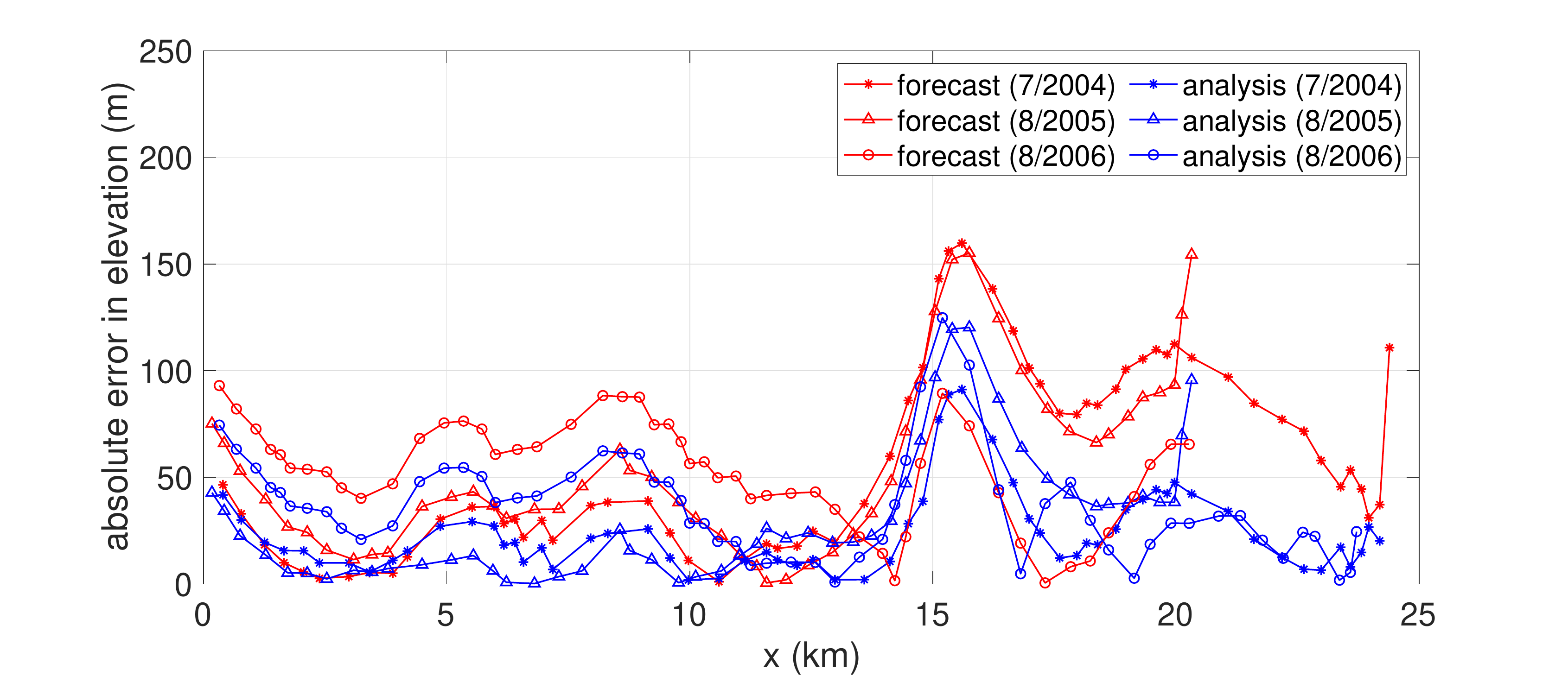}
    \caption{Absolute error in elevation for the forecast
      and analysis ensemble means in Experiment~3.}
    \label{fig:nick_abs_error}
  \end{center}
\end{figure} 

We close by computing the absolute errors in ice sheet elevation for the
simulated forecast and analysis means, compared with the observed
surface profiles from Fig.~\ref{fig:helheim_real_data}. These two errors
are plotted in Fig.~\ref{fig:nick_abs_error} for each analysis time,
which shows that the analysis means are more accurate than those for the
forecast. Also, the errors within the downward sloping (linear) portion
of the bedrock are relatively smaller than those within the non-linear
``humped'' portion.
%On the other hand, the Aug~2006 errors in both forecast
%and analysis are significantly larger than that for the other two
%observation times, primarily because of the very rapid advance
%experienced in 2006 due to the unusually cold weather that year.

%%%%%%%%%%%%%%%%%%%%%%%%%%%%%%%%%%%%%%%%%%%%%%%%%%%%%%%%%%%%%%%%%%
%%%%%%%%%%%%%%%%%%%%%%%%%%%%%%%%%%%%%%%%%%%%%%%%%%%%%%%%%%%%%%%%%%
\section{Conclusion}
\label{sec:conclusion}

We developed an algorithm that integrates an Ensemble Transform Kalman
Filter (ETKF) data assimilation scheme with an ice-flow model that
exploits the level set method to track the ice surface.  Accurate
simulations of ice sheet-ice stream systems are normally thought to
require use of higher order models~\cite{kirchner-etal-2011}. However,
our hybrid SSA--LSM--ETKF algorithm, using a zeroth-order SSA model, takes special care to accurately capture the terminus position with a LSM and
assimilates observational data for both ice surface elevation and
terminus position to improve solution fidelity.
%Our hybrid SSA--LSM--ETKF algorithm assimilates
%observational data for both ice surface elevation and terminus position.  
Based on two idealized marine-terminating glacier experiments and a simulation
of the retreat-advance cycle of Helheim Glacier in southeast
Greenland, we demonstrate 
that our data assimilation approach can seamlessly track seasonal and
multi-year variability in terminus position and ice surface
elevation. This model successfully tracks the ice surface elevation and
terminus positions during advance and retreat cycles.

We emphasize that our level set approach allows the terminus position
and ice surface to be tracked accurately and efficiently, despite using
a simple finite difference discretization on a fixed grid. The LSM
approach is robust and should be relatively easy to couple with any ice
sheet model from the shallow shelf approximation to a full Stokes flow
solver. And even though this paper is restricted to 2D model geometries,
it would be straightforward to extend the LSM algorithm to 3D geometries
having a dynamically evolving 2D glacier surface and a terminus
line. Our data assimilation approach has the additional advantage of
being a useful platform for parameter estimation.  More specifically,
when good estimates are unavailable for model parameters (such as melt
rate, front stress perturbation, etc.) we can simply choose a range for
each unknown parameter and distribute those values over ensembles,
taking advantage of observed data to pinpoint an ``optimal'' ensemble
member. Results of model simulations for experiments on idealized
marine-terminating glaciers and Helheim Glacier are encouraging and
suggest that this model could be exploited to improve the predictive
capability of ice sheet modelling.  Marine outlet glaciers (such as
Helheim) experience dynamic instabilities due to sensitivity to bed
topography and perturbations occurring at ice--atmosphere, ice--ocean,
and ice--bed boundaries. The nonlinear response of these glaciers to
climate forcing has profound implications for making projections of
future mass loss from ice sheets and the resulting sea-level rise.
Therefore, numerical techniques such as the one described here could
play an important role in efforts to make more accurate predictions of
ice mass loss due to climate change.

%%%%%%%%%%%%%%%%%%%%%%%%%%%%%%%%%%%%%%%%%%%%%%%%%%%%%%%%%%%%%%%%%%
\section*{Acknowledgments}

This work was supported by Discovery Grants (to S.P. and J.M.S.) from
the Natural Sciences and Engineering Research Council of Canada (NSERC).

\bibliographystyle{abbrv}  % alpha-order
\bibliography{refs_jcompV13}

% \bibliography{../../bib/ref,../../bib/lsm,../../bib/da,../../bib/glaciology,../../bib/igsrefs}

% \section*{Supplementary Material}
% Supplementary material that may be helpful in the review process should

\end{document}